\documentstyle[12pt,a4]{article}

\makeatletter
\def\eqalign#1{\null\,\vcenter{\openup\jot\m@th
  \ialign{\strut\hfil$\displaystyle{##}$&$\displaystyle{{}##}$\hfil
      \crcr#1\crcr}}\,}
\def\eqalignleft#1{\null\,\vcenter{\openup\jot\m@th
  \ialign{\strut$\displaystyle{##}$\hfil&$\displaystyle{{}##}$\hfil
      \crcr#1\crcr}}\,}
\makeatother

\begin{document}

\title{Impossible solutions?}

 \author{A. D. Popova}
 \date{October 18, 2000}
 \maketitle
 \begin{center}
 {\em Sternberg Astronomical Institute, Moscow, Russia, E-mail:
 popova@sai.msu.ru}
 \end{center}

\begin{abstract}
We present $n$-dimensional vortex-ring-like and potential-like
solutions with unusual properties related to some elliptical
differential equations with compact sources. Solutions have almost 3-
or 2-dimensional behaviour in the spaces with arbitrary (large) odd and
even dimensions, respectively.
\end{abstract}

\section{Introduction}

It is well-known that in the n-dimensional ($n$D) space the
gravitational and Coulomb-like electric potentials possess the fall-off
laws $\sim R^{-(n-2)}$ for $n>2$, $R$ being a distance from a compact
source, i.e., the source located in a confined space domain. In the
case $n=2$, the law has the form $\sim \ln(1/R)$. This basic statement
is undoubtedly true. However, we can pose the question: Is it possible
that in the $n$D space some compact sources exist for which the
fall-off law of a potential-like physical quantity is adequate to the
law associated with the 3D or 2D space? - We give a positive answer to
this question. The 3D or 2D behaviour of some quantities takes place in
spaces with an arbitrary odd or even number of dimensions,
respectively; moreover, we mean not only a far-zone asymptotics, but a
near-zone one as well.

A situation can be explained as follows. Let us imagine that we split
the $n$D space onto 2D and $(n-2)$D subspaces. Consider a function -- a
geometrical object -- with a complicated transformation properties.  It
does not matter whether it is a scalar, or a vector orthogonal to the
radial direction in the 2D space. Let it simultaneously be a component
of some polyvector in the $(n-2)$D space. The covariant operator
$\nabla^2$ applied to such the function is not obliged to be an obvious
scalar Laplace operator. It can has another appearance, and we call
some operators, we dealt with, anti-Laplacians.  And the solution of
the corresponding Poisson-like equation with a compact source is not
obliged to be $\propto R^{-(n-2)}$. In the 3D space such a situation is
hardly distinguishable. Indeed, the remaining single dimension (denote
it by $z$) in the $(3-2)$D space provides the term $\partial^2/\partial
z^2$ in $\nabla^2$, the same for a 1-vector and for a scalar, leading
to the solution $\propto R^{-1}$ in both the cases.

A physically meaningful aspect of this situation is that given such the
fall-off laws, we cannot determine an exact dimensionality of our
physical space. We can only extract a "best-fit" value $n=2$ or $n=3$
from observations and only state that the real dimensionality is odd or
even. Thus, the effective dimensionality of our space can be dynamical
in essence. Perhaps, such compact sources are just elementary
particles, so that we have no contradictions with the physical picture
of our world.

The author's intention here is not to deal with any special physical
models. It seems to be enough to describe several mathematical facts
which weaken the mentioned basic statement. In this paper, we undertake
the systematic study of an outlined subject, which was originated from
considering vortex rings: The notion of the anti-Laplacians appeared
from the latter. The logic of our account repeats that of developing
this study.

In Sec.2, we recall the reader how to obtain a current function
describing an infinitely thin vortex ring. A traditional derivation of
this subject in old and lovely manuals like \cite{Lamb,Appel} has a
drawback from the modern viewpoint: There is no appellation to $\delta$
functions, although the latter being quite convenient for a researcher.
Indeed, the vortex rings had been known well before P.A.M. Dirac
invented his $\delta$ function. We modify the traditional derivation by
including the $\delta$ functions. During this derivation we, for the
first time, meet an operator belonging to the type of operators which we call
anti-Laplacians.

In Sec.3, we consider an extension of equations suitable for vortex
rings onto $n$D space. On the l.h.s. of these equations anti-Laplacian
operators appear, however, we do not know a priori explicit forms of
their r.h.s., i.e., those of ring sources. In this section we only give
a constructive way of obtaining some solutions to homogeneous
equations. The solutions are finite everywhere except of a ring set of
points, and they could serve as those for the $n$D vortex rings. It
turns out that the cases of odd and even $n$ are principally
distinguished, and so are their asymptotics. The odd-$n$D solutions
have asymptotics of the 3D ones, whereas the even-$n$D solutions,
omitting some details, have that of 2D ones. It only remains to learn
explicit forms of the ring-like sources. However, in order to do it, we
have to study some properties of anti-Laplacians (and Laplacians as
well) in the next section.

Thus, we return in Sec.4 to the familiar Laplacians and Poisson
equations with $\delta$ sources possessing some symmetries in relevant
coordinate frames. [In Sec.4, we speak about the standard solutions
($\propto R^{-(n-2)}$) only.] We have found some transformation of an
$(n-2)$D solution into a $n$D one.  Moreover, the same transformation,
from $(n-2)$D to $n$D, is suitable for both the l.h.s. and r.h.s. of
the Poisson equations. Furthermore, the transformation is the same for
both the odd-$n$D and even-$n$D cases. Thus, for symmetries considered,
we can construct any $n$D solution if we know the above transformation
and the 3D and 2D solutions (or the 4D solution in a special symmetry).
In Sec.4, we also show that the extraction of an $n$D $\delta$ source
from the $n$D solution gives the same result, which can be obtained via
applying our transformation to the 3D or 4D sources.

Sec.5 is an independent study without referring to vortex rings. We
examine connections between Laplacians and anti-Laplacians which
provide a possibility of finding (our nonstandard) solutions to
Poisson-like equations similarly to Sec.2. We pay attention to one of
the anti-Laplacians which surprisingly leads to potential-like $n$D
solutions related to a one-point source. They certainly satisfy
homogeneous equations everywhere except this point. It is remarkable
that these solutions are simplified versions of those in Sec.3 and have
the same properties. It is also remarkable that we also find (another)
transformation from $(n-2)$D quantities to $n$D ones suitable both for
solutions and equations.  Thus, we construct a transformation machinery
which permits one to resolve the problem of finding points sources.

This is done in Sec.6. We obtain two different answers for the cases of
odd-$n$ and even-$n$ dimensions. In the former case the source is an
obvious $\delta$-like one, however, in the latter case the situation is
more intricate. In order to input the transformation machinery, we
oblige to have in hand the case $n=4$. It is simple enough to be
calculated immediately, and we consider it in detail. The answer is
that the point-like source is not a $\delta$ source, which contradicts,
at first glance, to the known theorem. However, there are no true
contradiction; some discussion on this subject is also given.

In Sec.7., we return to vortex-ring-like solutions of Sec.3 and exhibit
their sources. We also describe some other ring-like solutions and give
several concluding remarks. Two appendices with necessary mathematical
information accomplish our work.

\section{Recalling a 3D vortex ring}

In the cylindrical coordinate frame $(r,\varphi, z)$, consider an
axially symmetric motion of incompressible fluid with the symmetry axis
directed along $z$, $r$ being the distance from $z$, and there is no
dependence of $\varphi$ (see \cite{Lamb, Lav-Shab, M-T}  or any other
suitable manuals). Let $V_r(r,z)$ and $V_z(r,z)$ be the $r$ and $z$
components of a fluid velocity field $\vec V$, respectively. Let the
continuity equation be satisfied everywhere, perhaps except some
points:
$$
\vec \nabla \cdot \vec V = {\partial \over \partial r} (rV_r) +
{\partial \over \partial z} (rV_z) = 0.
\eqno(2.1)
$$
In our symmetry assumptions, the only $\varphi$ component of the $\vec
V$ curl survives, let it be a given value $\Omega_\varphi(r,z)$,
$$
(\vec \nabla \times \vec V)_\varphi = {\partial V_r\over \partial z}
 - {\partial V_z \over \partial r} = \Omega_\varphi .
\eqno(2.2)
$$

It follows from (2.1) that the expression
$$
r V_z dr - r V_r dz = d\psi
$$
is a local differential, where the function $\psi(r,z)$ is called the
Stokes current function\footnote{Our choice of signs in (2.3) coincides
with that of \cite{Lav-Shab} and is opposite to that of \cite{Lamb,
M-T}.}, and the previous definition of $\psi$ is equivalent to
$$
V_r = -\>{1\over r}\> {\partial \psi \over \partial z}\>, \qquad
V_z = {1\over r}\> {\partial \psi \over \partial r}\>.
\eqno(2.3)
$$
Hence, equation (2.2) can be represented as that for $\psi$:
$$
\overline{\Delta}_r \psi \equiv r\>{\partial \over \partial r}\>{1\over r}
\>{\partial \psi \over \partial r} + {\partial^2 \psi \over \partial z^2}
= -r \Omega_\varphi (r,z).
\eqno(2.4)
$$
The operator defined on the l.h.s. of (2.4) is not a Laplace
operator.  In keeping in mind further generalizations, we call it
anti-$r$-Laplacian (we refer the reader to Sec.5).

In fact, the function $\psi$ makes the sense of the $\varphi$ component
of a vector field $\vec A$.  The identically vanishing divergence (2.1)
means that $\vec V$ is a curl vector, i.e.,
$$
\vec V = \vec \nabla \times \vec A.
$$
In our assumptions about the vector $\vec V$, the only $A_\varphi$
component of $\vec A(r,z)$ is nonzero and
$$
V_r = -\>{1\over r}\> {\partial A_\varphi \over \partial z}\>, \qquad
V_z = {1\over r}\> {\partial A_\varphi \over \partial r}\>.
\eqno(2.5)
$$
The comparison of (2.3) and (2.5) allows us to impose $\psi =
A_\varphi$; \footnote{This have a consequence $\psi = r|\vec A|$, cf.
loc. cit. with our sign convention.} This fact helps one to solve
equation (2.4) for $\psi$ by connecting the operator
$\overline{\Delta}_r$ with the Laplacian, see below. Our account is
given following the familiar way of \cite{Appel, Lamb}.

In the Cartesian coordinate frame $x,y,z$, where $x=r\cos \varphi$ and
$y=r\sin \varphi$, we take, e.g., the $y$ component of $\vec A$:
$$
A_y = {\cos\varphi \over r}\>A_\varphi
$$
(we could choose the component $A_x$ that would give the same final
result). In the above frame, the vector operator $\nabla^2$ applied to
any component of $\vec A$ coincides with the scalar operator, denoted
by $\Delta_{Cart}$, applied to the same component,\footnote{We denote
by $\nabla^2$ an entirely covariant operator acting, e.g., on a vector.
The notation $\Delta$ is reserved for an operator acting on a scalar
(e.g., $\Phi$) only: $\nabla^2\Phi\equiv \Delta\Phi$.} that is why
$$
\nabla^2 A_y = \Delta_{Cart}A_y \equiv
\left({\partial^2 \over \partial x^2} +
{\partial^2 \over \partial y^2 } + {\partial^2 \over \partial z^2}
\right) A_y =
$$
$$
\left({1\over r}\> {\partial \over \partial r}\>r\>{\partial \over \partial r}
+ {1\over r^2}\>{\partial^2 \over \partial \varphi^2} +
{\partial^2 \over \partial z^2} \right){\cos\varphi \over r}\>A_\varphi \equiv
\Delta_{(r,\varphi,z)}{\cos\varphi \over r}\>A_\varphi.
$$
Then, in replacing back $A_\varphi$ by $\psi$, it is easy to establish
the validity of the rearrangement
$$
\Delta_{(r,\varphi,z)}{\cos\varphi \over r}\>\psi =
{\cos\varphi \over r}\>\overline{\Delta}_r \psi.
\eqno(2.6)
$$
We stress once more that equation (2.6) is an important step which
allows one to construct a solution for $\psi$ (or $A_\varphi$)
generated by a $\delta$-like source using the fundamental solution of
the Poisson equation.

Consider an infinitely thin vortex ring of the radius $a$ but with the
finite vortex intensity $\kappa$: $\kappa=\Omega_\varphi dS$ where $dS$
is the square of the infinitesimal ring cross section. We should impose
$$
\Omega_\varphi = \kappa a\>{\delta (r-a)\over r}\>\delta (z)
$$
in order for the integral of $\Omega_\varphi$ over the whole volume to
be equal to $2\pi a dS\Omega_\varphi= 2\pi \kappa a$, leading to
equation (2.4) in the form
$$
\overline{\Delta}_r \psi = -\kappa a\>\delta (r-a) \delta (z).
\eqno(2.7)
$$
From (2.6) and (2.7), we obtain the Poisson-like equation:
$$
\Delta_{(r,\varphi,z)}\>{\cos\varphi \over r}\>\psi =
- \kappa a\>{\cos\varphi \over r}\>  \delta (r-a) \delta (z).
\eqno(2.8)
$$
Due to the compact character of the source on the r.h.s. of (2.8) the
solution can be found by a standard way:
$$
{\cos\varphi \over r}\>\psi = {\kappa a\over 4\pi}\>
\int\limits_0^{\infty} dr'\>r' \int\limits_0^{2\pi} d\varphi'
\int\limits_{-\infty}^{\infty} dz'\>{\cos\varphi' \over r'}\>
{\delta (r'-a)\>\delta (z')\over |\vec R -\vec R'|}
\eqno(2.9)
$$
where
$$
|\vec R -\vec R'|= [r^2 + r'^2 -
2r'r\cos (\varphi-\varphi') + (z-z')^2]^{1/2}
$$
is the distance between the points with the coordinates $(r,\varphi,z)$
and $(r',\varphi',z')$. Integrating (2.9) with respect to $r'$ and $z'$
gives
$$
{\cos\varphi \over r}\>\psi = {\kappa a\over 4\pi}
\int\limits_0^{2\pi} d\varphi'{\cos\varphi' \over [r^2 + a^2 -
2ar\cos (\varphi-\varphi') + z^2]^{1/2}}\>.
\eqno(2.10)
$$
In order to remove the factor $\cos\varphi$ on the l.h.s. of (2.10),
we displace the origin of
$\varphi'$: $\varphi'=\varphi+\alpha$, so that integration will be done
with respect to $\alpha$. After using the equality
$$
\cos \varphi'= \cos\varphi \,\cos\alpha - \sin\varphi \,\sin\alpha
$$
we ensure that
$$
\int\limits_0^{2\pi} d\alpha \>
{\sin\alpha \over [r^2 + a^2 -
2ar\cos\alpha + z^2]^{1/2}} = 0,
$$
and the above factor on the l.h.s. of (2.10) cancels with that on a
r.h.s. of a final expression.

The required solution for the 3D vortex ring can now be written in the
form
$$
\psi = {\kappa a\over 2\pi}\>r \int\limits_0^{\pi} d\alpha \>
{\cos \alpha \over [r^2 + a^2 -
2ar\cos\alpha + z^2]^{1/2}}\>.
\eqno(2.11)
$$
And here we stop as yet. Our task was to only obtain (2.11) for
comparison with further extensions on multidimensional spaces. It is
well known how to express $\psi$ via elliptical functions, there are
also known its various properties and asymptotics, see, e.g., already
cited manuals and many others. Note that instead of the vortex ring, we
could speak about a ring electric contour, $A_{\varphi}(=\psi)$ being
the $\varphi$ component of the electromagnetic vector potential. --
There are no difference from the mathematical viewpoint.

\section{The $n$D vortex-ring-like solutions}

In order to describe an $n$-dimensional ring structure, which is a
direct product of the ring of the radius $a$ on an infinitesimal
$(n-1)$-dimensional ball, it is convenient to introduce an $(r,z)$
coordinate frame corresponding to the $2+(n-2)$ splitting of
$R^n$: $R^n=R^2\times R^{n-2}$. In each space, $R^2$ and $R^{n-2}$, we
construct spherical coordinate frames with the radial coordinates $r$
and $z$, and the angle coordinates $\varphi$ and
$\theta_1,\theta_2,\ldots,\theta_{n-3}$, respectively (see Appendix A).

From now we shall not confine ourselves by any physical interpretation.
Consider a vector field $\vec v(r,z)$ which has the only $v_r(r,z)$ and
$v_z(r,z)$ nonvanishing components. Let the divergence of this vector
field be equal to zero almost everywhere,
$$
\vec \nabla \cdot \vec v = {1\over r}\>{\partial \over \partial r}\>
(rv_r) + {1\over z^{n-3}}\>{\partial \over \partial z}\>(z^{n-3}v_z)=0
\eqno(3.1)
$$
[see (A.5), also note that in the frame selected $v^z=v_z$, $v^r=v_r$].
If we impose, as before in Sec.2,
$$
v_r= -\>{1\over rz^{n-3}}\>{\partial \psi \over \partial z}\>, \qquad
v_z= {1\over rz^{n-3}}\>{\partial \psi \over \partial r}\>,
\eqno(3.2)
$$
then the divergence (3.1) vanishes identically.

The extension of the curl operator to the $n$D space leads to a
covariant 2-vector or a contravariant $(n-2)$-vector, see, e.g.,
\cite{Schouten}. However, this fact does not prevent us to propose
that its only nonvanishing covariant component is unknown
as yet value $\omega_{r,z}$:
$$
{\partial v_r\over \partial z} - {\partial v_z\over \partial r}=
\omega_{r,z}.
\eqno(3.3)
$$
After substituting (3.2) into (3.3), an operator can be defined which
we call anti-double-Laplacian and denote by $\overline{\overline{\Delta}}
{}^{(n)}$:
$$
\overline{\overline{\Delta}}{}^{(n)}\psi_n = \left( r\>
{\partial \over \partial r}\> {1\over r}\>
{\partial \over \partial r} + z^{n-3}\> {\partial \over \partial z}\>
{1\over z^{n-3}}\>{\partial \over \partial z}\right) \psi_n
= - rz^{n-3}\omega_{r,z}.
\eqno(3.4)
$$

Now we shall find solutions to the homogeneous (anti-double-Laplace)
equations for all $n\ge 3$,
$$
\overline{\overline{\Delta}}{}^{(n)}\psi_n = 0,
\eqno(3.5)
$$
keeping in mind that $\omega_{r,z}$ should be considered equal to zero
almost everywhere, except of the set of points $r=a$, $z=0$, because we
suggest to search for the vortex-ring-like solutions. This is done below,
separately for odd and even $n$.

\subsection{The odd-$n$D solutions}

Consider the probe function
$$
\overline{\psi}_k = r \int\limits_0^{\pi}d\alpha\> \cos\alpha \>
{z^{k-3}\over R^{k-2}}
\eqno(3.6)
$$
where we denote
$$
R = \sqrt{\rho^2 + z^2},
\eqno(3.7a)
$$
$$
\rho^2 = r^2 + a^2 -2ar\cos\alpha .
\eqno(3.7b)
$$
In applying the operator $\overline{\overline{\Delta}}{}^{(n)}$
to $\overline{\psi}_k$, we come to the equality
$$
\overline{\overline{\Delta}}{}^{(n)}\overline{\psi}_k =
r \int\limits_0^{\pi}d\alpha\> \cos\alpha \left[(n-k)(k-2)\>
{z^{k-3}\over R^k} - (k-3)(n+1-k){z^{k-5}\over R^{k-2}}\right]
\eqno(3.8)
$$
after using the integral identity
$$
\int\limits_0^{\pi}d\alpha\>\cos\alpha
\left[{1\over R^{k-2}} -(k-2)\>{ar\cos\alpha\over R^k} +k(k-2)\>
{a^2r^2 \sin^2 \alpha\over R^{k+2}}\right] =
$$
$$
\int\limits_0^{\pi}d\alpha\>{\partial \over \partial \alpha}
\left[{\sin\alpha\over R^{k-2}} - (k-2)\>
{ar\sin\alpha \cos\alpha \over R^k}\right] = 0 .
\eqno(3.9)
$$
Let us explicitly write the formulae (3.8) for odd integer $k$ between
3 and $n$:
$$
\overline{\overline{\Delta}}{}^{(n)}\overline{\psi}_3 =
r \int\limits_0^{\pi}d\alpha\cos \alpha \>
(n-3)\>{1\over R^3},
\eqno(3.10a)
$$
$$
\overline{\overline{\Delta}}{}^{(n)}\overline{\psi}_5 =
r \int\limits_0^{\pi}d\alpha\cos \alpha \left[
- 2(n-4)\>{1\over R^3} +3(n-5){z^2\over R^5}\right],
\eqno(3.10b)
$$
$$
\ldots
$$
$$
\overline{\overline{\Delta}}{}^{(n)}\overline{\psi}_{n-2} =
r \int\limits_0^{\pi}d\alpha\cos \alpha \left[- 3(n-5)\>
{z^{n-7}\over R^{n-4}} + 2(n-4)\>{z^{n-5}\over R^{n-2}}
\right],
\eqno(3.10c)
$$
$$
\overline{\overline{\Delta}}{}^{(n)}\overline{\psi}_n =
r \int\limits_0^{\pi}d\alpha\cos \alpha \left[
-(n-3)\>{z^{n-5}\over R^{n-2}}\right].
\eqno(3.10d)
$$
We see that the terms with the factors $(k-3)$ and $(n-k)$ have
disappeared on the r.h.s. of (3.10a) and (3.10d) for $k=3$ and $k=n$,
respectively. The remaining terms can be compensated when taking
suitable combinations of $\overline{\psi}_k$: The single term ($\propto
1/R^3$) in (3.10a) can be compensated by the first term in (3.10b), and
so on, the last ($\propto z^{n-5}/R^{n-2}$) term in (3.10c) can be
finally compensated by the single term in (3.10d). Thus, the desired
solution can be constructed as a finite series of $\overline{\psi}_k$,
$$
\psi_n = \left.\sum_{k=3}^n \right.' a_{k,n}\overline{\psi}_k
= r\int\limits_0^{\pi}d\alpha\cos \alpha
\left.\sum_{k=3}^n \right.' a_{k,n}\>{z^{k-3}\over R^{k-2}}\>.
\eqno(3.11)
$$
where the prime over the sum indicates that the latter is
taken with respect to the only odd $k$. We omit henceforth any physically
meaningful coefficients. As to the coefficients $a_{k,n}$,
the recurrence relation takes place:
$$
a_{k+2,n} = a_{k,n}\>{(k-2)(n-k)\over (k-1)(n-k+1)}\>.
\eqno(3.12)
$$
Solving (3.12) with the (arbitrary) choice $a_{3,n}=1$ for every $n$
yields
$$
a_{k,n} = {1\cdot 3\cdot 5\cdots (k-4)(n-3)(n-5)\cdots (n-k+2)
\over 2\cdot 4\cdot 6\cdots (k-3)(n-4)(n-6)\cdots(n-k+1)}\>.
\eqno(3.13)
$$
More information about the properties of the coefficients $a_{k,n}$
is given in Appendix B.

In addition to already given solution (2.11) for $n=3$, we expose here
the three subsequent solutions (3.11) for $n=5,7,9$:
$$
\eqalign{
 \psi_3 & = \> r\int\limits_0^{\pi}d\alpha\cos \alpha \>{1\over R}\>,\cr
\psi_5 & = \> r\int\limits_0^{\pi}d\alpha\cos \alpha
\left({1\over R} + {z^2\over R^3}\right),\cr
\psi_7 & = \> r\int\limits_0^{\pi}d\alpha\cos \alpha \left(
{1\over R} + {2\over 3}\>{z^2\over R^3} + {z^4\over R^5}\right),\cr
\psi_9 & = \> r\int\limits_0^{\pi}d\alpha\cos \alpha \left(
{1\over R} + {3\over 5}\>{z^2\over R^3} + {3\over 5}\>{z^4\over R^5}
+ {z^6\over R^7}\right).\cr
}
$$

The far-zone asymptotics of the obtained solutions is very interesting
from the physical viewpoint. When $r\to \infty$ for any finite $z$, the
term $\overline{\psi}_k$ falls off as $\sim r^{2-k/2}$, so that for
every odd-$n$D solution independently of $n$ the term with $k=3$,
$1/R$, is asymptotically a leading one.  When $z\to \infty$ for any
finite $r$, all the terms $\overline{\psi}_k$ have the same asymptotics
$\sim z^{-1}$.  Thus, far from a compact source, $\psi_n$ behaves
almost like the 3D solution.

As for the near-zone asymptotics, the common feature of these solutions
is that for $z=0$ they exhibit the true 3D behaviour, and for every
finite $z$ the integrand of any $\psi_n$ is $\sim z^{-1}$ when $r\to 0$.

\subsection{The even-$n$D solutions}

It is clear from the consideration in the previous subsection that any
function of the (3.6)-kind is not suitable for the even-D case, hence
we probe other functions. For even integers $l>4$, we define
$$
\overline{\psi}_l = r\int\limits_0^{\pi}d\alpha\cos\alpha\>
{z^{l-4}\over R^{l-4}}\>,
\eqno(3.14)
$$
whereas for $l=4$, we give a distinct definition
$$
\overline{\psi}_4 = r\int\limits_0^{\pi}d\alpha\cos\alpha\>\ln {1\over R}\>.
\eqno(3.15)
$$
We can find that, for $l>4$
$$
\overline{\overline{\Delta}}{}^{(n)}\overline{\psi}_l =
r\int\limits_0^{\pi}d\alpha\cos \alpha \>(l-2)\left[(n-2-l)\>
{z^{l-2}\over R^l} - (n-l)\>{z^{l-4}\over R^{l-2}}\right],
\eqno(3.16)
$$
after using the integral identity (3.9) with $k$ replaced
by $l$, and for $l=4$
$$
\overline{\overline{\Delta}}{}^{(n)}\overline{\psi}_4 =
r\int\limits_0^{\pi}d\alpha\cos \alpha
\left[-(n-4)\>{1\over R^2}\right]
\eqno(3.17)
$$
after using the following integral identity
$$
\int\limits_0^{\pi}d\alpha \>{\cos\alpha\over R^4}\>
\left(\ln {1\over R} - {ar\cos\alpha\over R^2} + 2\>
{a^2r^2 \sin^2 \alpha\over R^4}\right) =
$$
$$
\int\limits_0^{\pi}d\alpha \>{\partial \over \partial \alpha}
\left( \sin\alpha \ln {1\over R} -
{ar \sin\alpha \cos\alpha \over R^2}\right) = 0.
\eqno(3.18)
$$
First of all, note that $\psi_4 =\overline{\psi}_4$ is a solution to
(3.5) for $n=4$, which is clear from (3.17):
$\overline{\overline{\Delta}}{}^{(4)} \psi_4=0$. As before, we also
explicitly write (3.16):
$$
\overline{\overline{\Delta}}{}^{(n)}\overline{\psi}_6 =
r\int\limits_0^{\pi}d\alpha\cos \alpha\>
2\left[(n-4)\>{1\over R^2} - (n-6)\>{z^2\over R^4}\right],
\eqno(3.19a)
$$
$$
\ldots
$$
$$
\overline{\overline{\Delta}}{}^{(n)}\overline{\psi}_{n-2} =
r\int\limits_0^{\pi}d\alpha\cos \alpha\>
(n-6)\left[4\>{z^{n-8}\over R^{n-6}} - 2\>{z^{n-6}\over R^{n-4}}
\right],
\eqno(3.19b)
$$
$$
\overline{\overline{\Delta}}{}^{(n)}\overline{\psi}_n =
r\int\limits_0^{\pi}d\alpha\cos \alpha\>
2(n-4)\>{z^{n-6}\over R^{n-4}}\>.
\eqno(3.19c)
$$
For $n\ne 4$, the appearance of (3.17) and (3.19) also demonstrates that
some linear combination of $\overline{\psi}_l$ can give the solution to (3.5):
The term $\sim 1/R^2$ in (3.17) can be compensated by the first term in
(3.19a), and so on, the last term in (3.19b) being compensated by the
single term in (3.19c). The solution can be written in the form
$$
\psi_n = a_{4,n}\overline{\psi}_4 + \left.\sum_{l=6}^n \right.''
a_{l,n} \overline{\psi}_l
\eqno(3.20)
$$
where the double prime means that the sum is taken with respect to the
only even $l$, and there is a simple recurrence relation for $a_{l,n}$
when $l\ge 6$:
$$
a_{l+2,n}=a_{l,n}\>{l-4\over l-2}\>.
\eqno(3.21)
$$
If we choose $a_{4,n}=1$, then $a_{6,n}=1/2$ from (3.17) and (3.19a),
and (3.21) can be easily resolved for $l\ge 6$:
$$
a_{l,n}={1\over l-4}\>.
\eqno(3.22)
$$
As a result of (3.20) and (3.22), the function
$$
\psi_n = r\int\limits_0^{\pi}d\alpha\cos\alpha
\left(\ln {1\over R} + \left.\sum_{l=6}^n \right.''{1\over l-4}
\>{z^{n-4}\over R^{n-4}}\right)
\eqno(3.23)
$$
for $n>4$ is the solution we just searched for.

Here it is pertinent to consider the most degenerate case $n=2$ where
the "$z$ part" of the operator (3.4) is absent, that is
$$
\overline{\overline{\Delta}}{}^{(2)}\psi_2 =
\overline{\Delta}{}_r^{(2)}\psi_2 = r\>{\partial \over \partial r}\>
{1\over r}\>{\partial \over \partial r}\> \psi_2 =0.
\eqno(3.24)
$$
As expected, the function
$$
\psi_2 = r\int\limits_0^{\pi}d\alpha\cos\alpha\>\ln {1\over \rho}\>.
\eqno(3.25)
$$
is a vortex-ring-like solution to (3.24) with $\rho$ defined by (3.7b);
during this derivation the integral identity similar to (3.18) is used:
$$
\int\limits_0^{\pi}d\alpha \cos\alpha
\left(\ln {1\over \rho} - {ar\cos\alpha\over \rho^2} + 2\>
 {a^2r^2\sin^2 \alpha\over \rho^4}\right) =
$$
$$
\int\limits_0^{\pi}d\alpha\>{\partial \over \partial \alpha}
\left( \sin\alpha \ln {1\over \rho} -
{ar \sin\alpha \cos\alpha \over \rho^2}\right) = 0.
$$

From the physical viewpoint, the solution (3.25) can be hardly
interpreted as a proper vortex ring because there is no an orthogonal
direction ($z$) to provide rotation around the ring axis. Nevertheless,
it is worth including in our collection because it is required for a study
described in Sec.7.

The distinct feature of the solution (3.23) is that the coefficients
$a_{l,n}$ do not involve $n$, so that the $(n+2)$D solution merely
acquires an extra term as compared with the $n$D one. For example,
the 4D solution and the three subsequent solutions with $l\ge 6$ are
as follows
$$
\eqalign{
 \psi_4 & = r\int\limits_0^{\pi}d\alpha\cos\alpha \ln {1\over R}\>, \cr
  \psi_6 & = r\int\limits_0^{\pi}d\alpha\cos\alpha
      \left(\ln {1\over R} + {1\over 2}\>{z^2\over R^2}\right), \cr
  \psi_8 & = r\int\limits_0^{\pi}d\alpha\cos\alpha
      \left(\ln {1\over R} + {1\over 2}\>{z^2\over R^2} +
      {1\over 4}\>{z^4\over R^4}\right), \cr
  \psi_{10} & = r\int\limits_0^{\pi}d\alpha\cos\alpha
      \left(\ln {1\over R} + {1\over 2}\>{z^2\over R^2} +
      {1\over 4}\>{z^4\over R^4} + {1\over 6}\>{z^6\over R^6}\right). \cr
}
$$

From the appearance of the solutions (3.15), (3.23) [and (3.25)], a far
asymptotics is easily seen because the term $\ln (1/R)$ is always
leading: When $r\to \infty$, the integrand of $\psi_n$ is always $\sim
\ln (1/r)$ starting from $n=2$ independently of $n$, and  when $z\to
\infty$, it is $\sim \ln (1/z)$ starting from $n=4$ independently of
$n$ as well.  This is just the 2D asymptotics.

In the hypersurface $z=0$, all these solutions behave like the 2D ones,
$\sim \ln (1/r)$; whereas for every finite $z$, the integrand of
$\psi_n$ is always $\sim \ln (1/z)+ Const$ when $r\to 0$.

\section{Known and unknown properties of Laplacians
and relevant quantities}
\subsection{The case of spherical symmetry}

It is widely known that the scalar Poisson equation in $n$
dimensions with the unit point source
$$
\Delta^{(n)}\Phi_n = - \sigma_n \delta (\vec R) \equiv s(\vec R)
\eqno(4.1)
$$
has solutions
$$
\Phi_n = {1\over n-2}\>{1\over |\vec R|^{n-2}}\>, \quad {\rm for}\quad n\ge 2,
$$
$$
\Phi_2 = \ln {1\over |\vec R|}\>, \quad {\rm for}\quad n=2,
$$
where $\vec R$ is the radius-vector originating from the point source, and
$$
\sigma_n={2\pi^{n/2}\over\Gamma(n/2)}
\eqno(4.2)
$$
with $\Gamma$ the gamma function is the surface of the
$(n-1)$-dimensional sphere of unit radius.  The expressions for
equation (4.1) and its solutions are indeed covariant, without referring
to any coordinate frame.

In the spherical coordinate frame (the $x$ frame), the entire
expression for the Laplacian involving all the $n-1$ angles is given in
Appendix A.  Consider the spherically symmetric case so that
$\Phi_n(\vec x) =\Phi_n(x)$ and the only radial dependence in (A.2)
survives. We also replace $\delta (\vec x)$ on $\delta(x)$ using the
formula\footnote{For accuracy, we add the index "+" to $\delta$,
because one often defines $\int_0^{\infty} dx\>\delta(x)\phi(x) =
\phi(0)/2$ for the singular point $x=0$ coinciding with a limit of
integration, see, e.g., \cite{Korn-Korn}.}

$$ \delta(\vec x) = {\delta_+(x)\over \sigma_n x^{n-1}}
\eqno(4.3)
$$
provided by the definition
$$ \int_{{\tau}^{(n)}}d\tau^{(n)} \delta(\vec x) \phi(x)=
\int\limits_0^{\infty} dx\>\delta_+(x)\phi(x) = \phi(0).
\eqno(4.4)
$$

Then, the truncated equation (4.1) acquires the form
$$
\Delta_x^{(n)}\Phi_n(x) \equiv
{1\over x^{n-1}}\> {\partial \over \partial x}\> x^{n-1} \>
{\partial \over \partial x}\>\Phi_n(x)= -\>{\delta_+(x)\over
x^{n-1}}\equiv s_x^{(n)}(x),
\eqno(4.5)
$$
having solutions
$$
\Phi_n(x)={1\over n-2}\>{1\over x^{n-2}}\>, \quad {\rm for}\quad n\ge 2,
\eqno(4.6a)
$$
$$
\Phi_2(x)=\ln {1\over x}\>, \quad {\rm for}\quad n=2.
\eqno(4.6b)
$$

Now consider a transformation which converts the $(n-2)$D solution
(4.6), $\Phi_{n-2}(x)$, into the $n$D one, $\Phi_n(x)$. For $n\ge 4$:
$$
\Phi_n(x)= -\>{1\over n-2}\>{1\over x}\>{\partial \over \partial x}\>
\Phi_{n-2}(x).
\eqno(4.7)
$$
The equality (4.7) is evident. We see that odd-$n$D solutions are
transformed into odd-$n$D ones starting from $n=5$, and even-$n$D
solutions are transformed into even-$n$D ones starting from $n=4$.  In
other words, every $n$D solution can be obtained in applying the above
transformation by required number of times to $\Phi_3$ or to $\Phi_2$.

Moreover, taking into account the following operator rearrangement
$$
{1\over x}\>{\partial \over                                                                        \partial x}\>\Delta_x^{(n-2)} =
\Delta_x^{(n)}\> {1\over x}\>{\partial \over \partial x}
\eqno(4.8)
$$
we obtain
$$
\Delta_x^{(n)}\Phi_n(x) =
 -\>{1\over n-2}\>{1\over x}\>{\partial \over \partial x}\>
\Delta_x^{(n-2)} \Phi_{n-2}(x).
\eqno(4.9)
$$
The equality (4.9) means that the l.h.s. of the Poisson equation (4.5)
is transformed similar to the solution itself.

Independently, the direct calculation with the explicit form of
$s^{(n-2)}(x)$ [see the definition on the r.h.s. of (4.5)] shows that
the similar situation takes place for sources:
$$
s_x^{(n)}(x)= -\>{1\over n-2}\>{1\over x}\>{\partial \over \partial x}\>
s_x^{(n-2)}(x)
\eqno(4.10)
$$
after using the formal equality $x{\delta'}_+(x)=-\delta_+(x)$
where the prime denotes differentiation.

\subsection{The case of double-spherical symmetry}

In the $(r,z)$ coordinate frame introduced before in Sec.3, the central
point source in (4.1) for the double-spherical symmetry can be expressed
as follows ($n>2$)
$$
s^{(n)}(\vec R) = -\sigma_n \delta(\vec r) \delta(\vec z) =
-\sigma_n \>{\delta_+(r)\over \sigma_2 r}\>
{\delta_+(z)\over \sigma_{n-2} z^{n-3}}= -\>{1\over n-2}\>
{\delta_+(r)\over r}\>{\delta_+(z)\over z^{n-3}}\>,
\eqno(4.11)
$$
where we have used (4.3), and the equality
$$
\sigma_n={\sigma_2\>\sigma_{n-2} \over n-2}
\eqno(4.12)
$$
which is easy to verify with referring to the definition (4.2),
also note that $\sigma_2 =2\pi$ and $\sigma_1=2$.

For the $(r,z)$ dependence of a solution, the truncated operator
(A.6) combined with the source (4.11) gives the Poisson-like
equation ($n>2$)
$$
\Delta_{r,z}^{(n)} \Xi_n(r,z) \equiv
\left( {1\over r}\> {\partial \over \partial r}\> r \>
{\partial \over \partial r} +
{1\over{z^{n-3}}}\> {\partial \over \partial z}\> z^{n-3} \>
{\partial \over \partial z} \right)
\Xi_n(r,z)= -\>{\delta_+(r)\over r} \> {\delta_+(z)\over z^{n-3}}
\equiv s_{r,z}^{(n)}
\eqno(4.13a)
$$
where the solution is
$$
\Xi_n(r,z)= {1\over(r^2+z^2)^{(n-2)/2}} \equiv
{1\over \bar R^{n-2}}\>.
\eqno(4.14a)
$$
Note that due to the remarkable formula (4.12), the factor
$(n-2)$ is removed both on the r.h.s. of (4.13a) and in the
solution (4.14a).

The case $n=2$ is degenerate because there are no the $z$ space,
and the Poisson equation
$$
\Delta_{r,z}^{(n)}\Xi_2(r) = \Delta^{(2)}\Xi_2(r) =
{1\over r}\> {\partial \over \partial r}\>r\>
{\partial \over \partial r}\>\Xi_2 = - \>{\delta_+(r)\over r}\>
\eqno(4.13b)
$$
has the trivial solution
$$
\Xi_2 = \ln {1\over r}\>.
\eqno(4.14b)
$$

In the case of the $(r,z)$ frame, we can also introduce a
transformation with properties similar to the above one despite of the
presence of the two variables, $r$ and $z$. It is easy to verify that
the required transformation is as follows ($n\ge 5$)
$$
\Xi_n(r,z)= -\>{1\over n-4}\>{1\over z}\>{\partial \over \partial z}\>
\Xi_{n-2}(r,z) \equiv f_L^{(n,n-2)}\Xi_{n-2}(r,z).
\eqno(4.15)
$$
The appearance of the factor $(n-4)$ instead of $(n-2)$ in (4.15) is
due to that the solutions (4.6a) and (4.14a) have distinct
coefficients. As before, the odd solutions are obtainable from the odd
ones starting from $n=5$. However unlike the previous case, our
transformation works starting from $n=6$ for even solutions. The point
is that, as mentioned, the $z$ dependence of the Laplace operator and
the corresponding solution disappears for $n=2$, see (4.14b), that is
why $\Xi_4$ cannot be obtained from $\Xi_2$.

Here, the rearrangement of the (4.8)-type also exists,
$$
{1\over z}\>{\partial \over \partial z}\>\Delta_{r,z}^{(n-2)} =
\Delta_{r,z}^{(n)}\> {1\over z}\>{\partial \over \partial z}\>,
$$
and leads to the equality analogous to (4.9):
$$
\Delta_{r,z}^{(n)} \Xi_n= f_L^{(n,n-2)}
\Delta_{r,z}^{(n-2)}\Xi_{n-2}.
\eqno(4.16)
$$
For the $\delta$-source in (4.13a), we also have
$$
s_{r,z}^{(n)}= f_L^{(n,n-2)}s_{r,z}^{(n-2)}.
\eqno(4.17)
$$
We call the set of quantities $\Xi_n$,
$\Delta_{r,z}^{(n)} \Xi_n$, and $s_{r,z}^{(n)}$ the Laplace set
$L_n$:
$$
L_n = \{\Xi_n,\>\Delta_{r,z}^{(n)} \Xi_n,\>s_{r,z}^{(n)}\},
\eqno(4.18)
$$
so that the equalities (4.15), (4.16) and (4.17) acquire the compact
notation
$$
L_n = f_L^{(n,n-2)} L_{n-2}.
\eqno(4.19)
$$
For odd and even $n$, any quantity from the Laplace set can be expressed as
follows
$$
L_n = f_L^{(n,n-2)} f_L^{(n-2,n-4)}\ldots f_L^{(5,3)} L_3 =
{1\over (n-4)!!} \left(-\>{1\over z}\>{\partial \over \partial z}
\right)^{(n-3)/2} L_3
\eqno(4.20)
$$
and
$$
L_n = f_L^{(n,n-2)}f_L^{(n-2,n-4)}\ldots f_L^{(6,4)} L_4 =
{1\over (n-4)!!} \left(-\>{1\over z}\>{\partial \over \partial z}
 \right)^{(n-4)/2} L_4,
\eqno(4.21)
$$
respectively.

\subsection{Methodological extraction of $\delta$-sources}

In methodological goals and for future applications in Sec.6, we show
how to prove that the $n$D solutions (4.14a) really provide corresponding
$\delta$ sources, i.e., to prove (4.13a). We shall demonstrate it by
two ways. The first of them can be found in any suitable manual in
functional analysis, see e.g., \cite{G-Sh}.  This is very trivial
derivation in the case of the $x$ frame, however it has some subtleties
in the case of the $(r,z)$ frame. The second way, just invented,
consists in using the introduced transformation $f_L^{(n,n-2)}$ in order
to construct a proof by a recurrent procedure.

Consider a function $\phi(r,z)\ni \cal D(G)$, where $\cal D(G)$ denotes
the class of trial functions, i.e. that of finite $C^{\infty}(\cal G)$
functions in the domain ${\cal G}(r,z)$: $supp\> \phi \subset {\cal G}(r,z)$.
Then,
$$
\int_{\tau^{(n)}} d\tau^{(n)} \phi\> \Delta_{r,z}^{(n)} \Xi_n =
\int_{\tau^{(n)}} d\tau^{(n)} (\Delta_{r,z}^{(n)}\phi)\>\Xi_n
\eqno(4.22)
$$
where $d\tau^{(n)}= d\sigma_2 d\sigma_{n-2}rz^{n-3}dr\>dz$, see
(A.4), and integration is performed over a large enough volume
$\tau^{(n)}$, so that the ranges of values of the $r$ and
$z$ coordinates contain ${\cal G}(r,z)$.

\vskip0.5cm

{\bf Traditional way}. Following \cite{G-Sh} and other manuals, we can write
$$\int_{\tau^{(n)}} d\tau^{(n)} (\Delta_{r,z}^{(n)}\phi)\>\Xi_n =
\lim_{\varepsilon,\eta \to 0} \int_{\tau_{\varepsilon,\eta}^{(n)}}
d\tau^{(n)} (\Delta_{r,z}^{(n)}\phi)\>\Xi_n
$$
$$
= \lim_{\varepsilon,\eta \to 0} \left( \int_{\tau_{\varepsilon,\eta}^{(n)}}
d\tau^{(n)}\phi\>\Delta_{r,z}^{(n)} \Xi_n + J_r +J_z\right)
\eqno(4.23)
$$
using the property (4.22), where the volume
$\tau_{\varepsilon,\eta}^{(n)}$ means that we exclude from $\tau^{(n)}$
the small domain near the point $r=0$, $z=0$;
$\tau_{\varepsilon,\eta}^{(n)}$: $r\ge \varepsilon$, $z\ge \eta$. The
quantities $J_r$ and $J_z$ denote the surface integrals:
$$
J_r = J_{r1} + J_{r2},
$$
$$
J_{r1} = \int_{S_r} dS_r \left[\phi\> {\partial \Xi_n \over \partial z}
\right]_{z=\eta} = - (n-2)\sigma_2 \sigma_{n-2} \eta^{n-2}
\int\limits_0^{\varepsilon} dr\>r\phi(r,\eta)\>
{1\over (r^2+\eta^2)^{n/2}}\>,
\eqno(4.24)
$$
$$
J_{r2} = - \int_{S_r} dS_r \left[{\partial \phi \over \partial z}\>\Xi_n
\right]_{z=\eta} = - \sigma_2 \sigma_{n-2} \eta^{n-2}
\int\limits_0^{\varepsilon} dr\>r\>\left.{\partial \phi \over \partial z}
\right|_{z=\eta}{1\over (r^2+\eta^2)^{(n-2)/2}}\>,
\eqno(4.25)
$$
and
$$
J_z = J_{z1} + J_{z2},
$$
$$
J_{z1} = \int_{S_z} dS_z \left[\phi\> {\partial \Xi_n \over \partial r}
\right]_{r=\varepsilon} = - (n-2)\sigma_2 \sigma_{n-2}\varepsilon^2
\int\limits_0^{\eta} dz\>z^{n-3}
\phi(\varepsilon,z)\>{1\over (\varepsilon^2+z^2)^{n/2}}\>,
\eqno(4.26)
$$
$$
J_{z2} =  - \int_{S_z} dS_z \left[{\partial \phi \over \partial r}\>\Xi_n
\right]_{r=\varepsilon} = - \sigma_2 \sigma_{n-2}\varepsilon^2
\int\limits_0^{\eta} dz\>z^{n-3}\left.{\partial \phi \over \partial r}
\right|_{r=\varepsilon} {1\over (\varepsilon^2+z^2)^{(n-2)/2}}\>,
\eqno(4.27)
$$
where
$$
dS_r = d\sigma_2 d\sigma_{n-2} dr\>r\eta^{n-3}
$$
and
$$
dS_z = d\sigma_2 d\sigma_{n-2} dz\>\varepsilon z^{n-3}
$$
are the $(n-1)$-dimensional surface elements. The 3-dimensional
analogue of $S_r$ is the base surface of the cylinder $0\le
r\le\varepsilon$, $-\eta\le z\le\eta$, and that of $S_z$ is the lateral
surface of the same cylinder. (Note that a $0$-dimensional sphere here
consists of the two isolated points $z=-\eta$ and $z=\eta$.)

In returning to (4.23), we note, first of all, that the prelimit term
involving $\Delta_{r,z}^{(n)} \Xi_n$ vanishes. Further, there are two
ways of making a limit procedure depending on what limit (with respect
to $\varepsilon$ or $\eta$) is taken first. From the expressions (4.24),
(4.25) and (4.26), (4.27) we see that always
$$
\lim_{\varepsilon \to 0} J_r = 0, \quad \lim_{\eta \to 0} J_z = 0
$$
meaning that when the order of taking limits in (4.23) is $\varepsilon
\to 0$ and then $\eta \to 0$ (or $\eta \to 0$, and then $\varepsilon
\to 0$) the only integral $J_z$ (or $J_r$) "works".

It can be easily shown that the following limit expressions of the
integrals containing the derivatives of $\phi$ vanish
$$
\lim_{\eta \to 0} J_{r2} = 0, \quad \lim_{\varepsilon \to 0} J_{z2} = 0.
$$
The calculation of the remaining integrals (4.24) and (4.26) will be
exhibited in more details.  The trivial integration in (4.24), mutual
for both odd-$n$D and even-$n$D cases, leads to
$$
\lim_{\eta\to 0}J_{r1} = - \sigma_2 \sigma_{n-2} \bar \phi_{\varepsilon}
$$
with $\bar \phi_{\varepsilon}$ some average value of $\phi(r,0)$ over the
interval $0\le r\le\varepsilon$.

As to the integral in (4.26), we should use different handbook integrals
for the odd-$n$D and even-$n$D cases. For $n=2q+1$, the integral in
(4.26) corresponds to the following one
$$
\int dz {z^{2(q-1)}\over (\varepsilon^2+z^2)^{q+1/2}} =
{z^{2q-1}\over (2q-1)\varepsilon^2(\varepsilon^2+z^2)^{q-1/2}}\>,
$$
see \cite{Pr-Br-M}, p.91, no.7. For even $n$ it is more convenient to
impose $n=2(m+2)$, $m=0$ corresponding to $n=4$, then the integral in
(4.26) can be represented as follows, see loc. cit., p.30, no.6,
$$
\int dz\>
{z^{n-3}\over (\varepsilon^2+z^2)^{n/2}} =
{1\over 2}\int du\>{u^m\over (\varepsilon^2+u)^{m+2}} =
-{1\over 2}\sum\limits_{k=0}^m {m\choose k}
{(-\varepsilon^2)^{m-k}\over (m-k+1)(\varepsilon^2+u)^{m-k+1}}
$$
where we have denoted $u=z^2$ and
$$
{m\choose k}= {m!\over k!(m-k)!}
$$
is the standard convention for a binomial coefficient. However, in both the
odd-$n$ and even-$n$ cases, we have the same limit expression
$$
\lim_{\varepsilon \to 0} J_{z1} = - \sigma_2 \sigma_{n-2}\bar \phi_{\eta}
$$
with $\bar \phi_{\eta}$ some average value of $\phi(0,z)$ over the
interval $0\le z\le\eta$. (For the even-$n$D case we have
required to additionally derive that
$$
\sum\limits_{k=0}^m {m\choose k} {(-1)^{m-k}\over m-k+1} =
{(-1)^m\over m+1}\left[\sum\limits_{k=0}^{m+1}(-1)^k {m+1\choose k}
- (-1)^{m+1}\right]={1\over m+1} = {2\over n-2},
$$
note that the first sum in the square brackets is zero, see loc. cit,
p.606, no.3.)

In returning to (4.23),
$$
\int d\tau^{(n)} \phi\> \Delta_{r,z}^{(n)} \Xi_n  =
\lim_{\varepsilon \to 0} \left(\lim_{\eta \to 0} J_{r1}\right) =
\lim_{\eta \to 0} \left(\lim_{\varepsilon \to 0} J_{z1}\right)
$$
$$
= - \sigma_2 \sigma_{n-2} \phi(0,0) = - (n-2) \sigma_n \phi(0,0),
\eqno(4.28)
$$
and our proof is finished. The appearance of the factor $(n-2)$ is due
to that it was removed in the solution (4.14a). Thus, (4.28) is equivalent
to equation (4.13a).

Below we shall use (4.28) in an equivalent form
$$
\int\limits_0^{\infty} dr\>r \int\limits_0^{\infty} dz\>z^{n-3} \phi\>
\Delta_{r,z}^{(n)} \Xi_n \equiv
\left< rz^{n-3} \phi,\> \Delta_{r,z}^{(n)} \Xi_n \right>
= - \phi(0,0).
\eqno(4.29)
$$

{\bf The recurrent way}. Let us propose that we have proved, perhaps
by the first method, that
$$
\Delta_{r,z}^{(3)} \Xi_3(r,z) = -\>{\delta_+(r)\over r}\> \delta_+(z)
$$
and
$$
\Delta_{r,z}^{(4)} \Xi_4(r,z) = -\>{\delta_+(r)\over r}\>
{\delta_+(z)\over z}\>,
$$
after that we can use the mathematical induction method. We prove
that if for $k=5,6,7,8,\ldots$,
$$
\Delta_{r,z}^{(k-2)} \Xi_{k-2}(r,z) = -\>{\delta_+(r)\over r} \>
{\delta_+(z)\over z^{k-5}}\>,
\eqno(4.30)
$$
then
$$
\Delta_{r,z}^{(k)} \Xi_{k}(r,z) = -\>{\delta_+(r)\over r} \>
{\delta_+(z)\over z^{k-3}}\>.
\eqno(4.31)
$$

Equation (4.30) means that [see (4.29)]
$$
\left< rz^{k-5} \phi,\> \Delta_{r,z}^{(k-2)} \Xi_{k-2} \right> = - \phi(0,0)
\eqno (4.32)
$$
for any $\phi(r,z)\ni \cal D(G)$. The required proof is given below by a
chain of relations with using the transformation $f_L^{(k,k-2)}$:
$$
\left< rz^{k-3} \phi,\> \Delta_{r,z}^{(k)} \Xi_k \right> =
\left<  rz^{k-3} \phi,\> f_L^{(k,k-2)}\Delta_{r,z}^{(k-2)} \Xi_{k-2}\right> .
\eqno(4.33)
$$
Here we recall that for every $\phi$ belonging to the mentioned class,
first, any derivatives of $\phi$ belong to the same class, and, second,
for any function $F(r,z)$ the equality
$$
\left< \phi,\>{\partial F\over \partial z} \right> =
- \left< F,\>{\partial \phi\over \partial z}\right> ,
$$
holds. Let us continue,
$$
{\rm (4.33)} = {1\over k-4} \left< r \>
{\partial \over \partial z}\>(\phi z^{k-4}),\>
\Delta_{r,z}^{(k-2)} \Xi_{k-2}\right> =
$$
$$
\left< r z^{k-5}\left(\phi
+ {z\over k-4}\>{\partial \phi\over \partial z}\right),\>
\Delta_{r,z}^{(k-2)} \Xi_{k-2}\right>  = - \phi(0,0).
$$
In the latter expression, the second term in parentheses vanishes when
integrated with $\Delta_{r,z}^{(k-2)} \Xi_{k-2}$. This is due to the
(formal) equality $z\delta_+(z)=0$ after using (4.30), or this fact can
be verified immediately. The first term there just gives (4.32). Thus,
(4.31) is proved.

\section{Anti-Laplacians: entirely unknown properties}
\subsection{Connections of anti-Laplacians with Laplacians}

Until now we have already introduced the anti-$r$-Laplacian and
anti-double-Laplacian operators. Here, we define the remaining
anti-Laplacians and derive some relations connecting any
anti-Laplacians with Laplacians. The way by which (2.6) was obtained
suggests that we propose a relation given below.

Consider for simplicity the $n$D space in the Cartesian coordinate
frame $x_1,x_2$, $\ldots$, $x_n$, let $x=\sqrt{x_1^2+x_2^2+\ldots
+x_n^2}$. We denote

$$
\Delta_{Cart}^{(n)}= {\partial^2 \over \partial x_1^2} +
{\partial^2 \over \partial x_2^2} + \ldots +
{\partial^2 \over \partial x_n^2}
$$
the $\Delta^{(n)}$ operator in the Cartesian frame. For any function $F$
depending only on $x$: $F=F(x)$, and for any coordinate $x_k$,
the following proposition holds:
$$
\Delta_{Cart}^{(n)}\left({x_k\over x^n}\>F(x)\right)
= {x_k\over x^n}\> \overline{\Delta}{}_x^{(n)} F(x)
\eqno(5.1)
$$
where the operator
$$
\overline{\Delta}{}_x^{(n)} \equiv
x^{n-1}\>{\partial \over \partial x}\>
{1\over x^{n-1}}\> {\partial \over \partial x} =
{\partial^2 \over \partial x^2} - {n-1\over x}\>
{\partial \over \partial x}
$$
should be called anti-Laplacian. (The reader has already understood
that the prefix "anti" is given because of the inverse position of
"$x$" and "$1/x$" in the first expression and the opposite sign in the
second expression as compared to the Laplacian).

In fact, Eq.(5.1) has a covariant meaning: it is independent of a
coordinate frame. We reformulate it in the $x$ frame. Let $\lambda_1$ be
a senior angle and $x_1$ be a Cartesian axis associated with
$\lambda_1$ (see Appendix A. We could also take an arbitrary axis $x_k$
involving other angles, however this means dealing with more cumbersome
expressions.) Then,
$$
{\Delta}_{x,\lambda_1}^{(n)}\left({\cos \lambda_1\over x^{n-1}}\>F(x)
\right)={\cos \lambda_1\over x^{n-1}}\>\overline{\Delta}{}_x^{(n)}F(x)
\eqno(5.2)
$$
where we have denoted
$$
{\Delta}_{x,\lambda_1}^{(n)}= {1\over x^{n-1}}\> {\partial \over \partial x}\>
x^{n-1}\>{\partial \over \partial x} +
{1\over x^2\>\sin^{n-2}\lambda_1}\>{\partial \over \partial \lambda_1}\>
\sin^{n-2}\lambda_1\>{\partial \over \partial \lambda_1}\>,
$$
the part of a full operator (A.2) acting on any function of $x$ and $\lambda_1$.

In the $(r,z)$ frame, the $n$D anti-$r$-Laplacian can be defined by
analogy with (2.4):
$$
\overline{\Delta}{}_r^{(n)}=r\> {\partial \over \partial r}\>
{1\over r}\>{\partial \over \partial r} +
{1\over{z^{n-3}}}\> {\partial \over \partial z}\> z^{n-3} \>
{\partial \over \partial z},
\eqno(5.3)
$$
and we also define the operator
$$
\overline{\Delta}{}_z^{(n)} \equiv  {1\over r}\>{\partial \over \partial r}\>
r\>{\partial \over \partial r} + z^{n-3}\> {\partial \over \partial z}\>
{1\over z^{n-3}}\>{\partial \over \partial z},
\eqno(5.4)
$$
which following our "taxonomy" is natural to call anti-$z$-Laplacian
and which plays a key role in our investigations. It accomplishes our
collection of anti-Laplacians for a given splitting of the $n$D space.

Given the $(r,z)$ frame, the relation (5.2) can be written
independently for the $r$ and/or $z$ spaces. For example, recall Eq.(2.6)
in 3 dimensions. In $n$ dimensions, for any $F(r,z)$, the analogue of (2.6) is
$$
{\Delta}_{r,\varphi,z}^{(n)}\left(
{\cos\varphi\over r}\> F(r,z)\right)
={\cos\varphi\over r}\>
\overline{\Delta}{}_r^{(n)}F(r,z),
\eqno(5.5)
$$
plus there are the relations in the $z$ space:
$$
{\Delta}_{r,z,\theta_1}^{(n)}\left(
{\cos \theta_1\over z^{n-3}}\> F(r,z)\right)
={\cos \theta_1\over z^{n-3}}\>
\overline{\Delta}{}_z^{(n)}F(r,z)
\eqno(5.6)
$$
and
$$
{\Delta}_{r,\varphi,z,\theta_1}^{(n)}\left({\cos\phi \over r}\>
{\cos \theta_1\over z^{n-3}}\> F(r,z)\right)
={\cos\varphi \over r}\> {\cos \theta_1\over z^{n-3}}\>
\overline{\overline{\Delta}}{}^{(n)}F(r,z)
\eqno(5.7)
$$
where we have denoted
$$
{\Delta}_{r,\varphi,z}^{(n)}= {1\over r}\> {\partial \over
\partial r}\> r\>{\partial \over \partial r} + {1\over z^{n-3}}\>
{\partial \over \partial z}\> z^{n-3}\>{\partial \over \partial z} +
{1\over r^2}\>{\partial^2 \over \partial \varphi^2}\>,
$$
$$
{\Delta}_{r,z,\theta_1}^{(n)}={1\over r}\> {\partial \over
\partial r}\> r\>{\partial \over \partial r} + {1\over z^{n-3}}\>
{\partial \over \partial z}\> z^{n-3}\>{\partial \over \partial z} +
{1\over z^2\>\sin^{n-4}\theta_1}\>{\partial \over \partial \theta_1}\>
\sin^{n-4}\theta_1\>{\partial \over \partial \theta_1},
$$
and
$$
{\Delta}_{r,\varphi,z,\theta_1}^{(n)} =
{\Delta}_{r,\varphi,z}^{(n)} +
{1\over z^2\>\sin^{n-4}\theta_1}\>{\partial \over \partial \theta_1}\>
\sin^{n-4}\theta_1\>{\partial \over \partial \theta_1},
$$
the parts of the expression (A.6) involving required variables.

Besides of the relations (5.5), (5.6) and (5.7), there exist some evident
differential relations. Let us turn again to the $x$ frame. Then, for
any function $F(x)$,
$$
x^{n-1}\>{\partial \over \partial x}\> \Delta_x^{(n)} F(x)=
\overline{\Delta}{}_x^{(n)} \left( x^{n-1}\>{\partial \over \partial x}\>
F(x) \right)
\eqno(5.8a)
$$
and
$$
{1\over x^{n-1}}\>{\partial \over \partial x}\> \overline{\Delta}{}_x^{(n)}
F(x)= \Delta_x^{(n)} \left({1\over x^{n-1}} \>{\partial \over \partial x}\>
F(x) \right).
\eqno(5.8b)
$$
The equalities (5.8) are in fact identities after using
explicit forms of operators. Thus, we have demonstrated a principle of
constructing such the relations.

In the $(r,z)$ frame, a variety of relations similar
to (5.8a) arises, based on the fact of commutativity of the
derivatives $\partial /\partial r$ and $\partial /\partial z$. For
any $F(r,z)$
$$
r\>{\partial \over \partial r}\> \Delta_{r,z}^{(n)} F(r,z)=
\overline{\Delta}{}_r^{(n)} \left( r\>{\partial \over \partial r}\>
F(r,z) \right),
\eqno(5.9a)
$$
$$
z^{n-3}\>{\partial \over \partial z}\> \Delta_{r,z}^{(n)} F(r,z)=
\overline{\Delta}{}_z^{(n)}\left( z^{n-3}\>{\partial \over \partial z}\>
F(r,z) \right),
\eqno(5.9b)
$$
$$
z^{n-3}\>{\partial \over \partial z}\>\overline{\Delta}{}_r^{(n)}F(r,z)=
\overline{\overline{\Delta}}{}^{(n)} \left( z^{n-3}\>
{\partial \over \partial z}\>F(r,z) \right).
\eqno(5.9c)
$$
There are no necessity to bring all the remaining relations which
correspond to (5.8b). We shall use below one of them, namely
$$
{1\over z^{n-3}}\>{\partial \over \partial z}\>
\overline{\Delta}{}_z^{(n)} F(r,z) = \Delta_{r,z}^{(n)}
\left({1\over z^{n-3}} \>{\partial \over \partial z}\>
F(r,z) \right).
\eqno(5.10)
$$

Moreover, we can in principle combine the relations of the type
(5.5)--(5.7) and those of the type (5.9) and (5.10).

\subsection{Potential-like solutions of homogeneous equations}

In keeping in mind that the point-like source is located at the point
$r=0,z=0$, it turns out that $n$D solutions to homogeneous equations
involving the operator (5.4), i.e., the anti-$z$-Laplace equations
$$
\overline{\Delta}{}_z^{(n)}\Psi_n \equiv  \left( {1\over r}\>
{\partial \over \partial r}\> r\>{\partial \over \partial r}
+ z^{n-3}\> {\partial \over \partial z}\>
{1\over z^{n-3}}\>{\partial \over \partial z}\right) \Psi_n = 0,
\eqno(5.11)
$$
are the simplified versions of the solutions (3.11) and (3.15), (3.23).
As before, we should consider separately the cases of odd and even
dimensions.

In the odd-$n$D case, the function
$$
\Psi_n = \left.\sum_{k=3}^{n}\right.' a_{k,n}\>{z^{k-3}\over \bar R^{k-2}}
\eqno(5.12)
$$
is the solution to (5.11) with $a_{k,n}$ given by (3.12) and (3.13),
recall that $\bar R=\sqrt{r^2+z^2}$. It is worth noting that the
recurrence relation (3.12) for $a_{k,n}$ is a necessary and sufficient
condition of the validity of (5.12), without any auxiliary construction
like the integral identity (3.9).

In the even-$n$D case, the functions
$$
\Psi_4 = \ln {1\over \bar R}\>,
\eqno(5.13a)
$$
and
$$
\Psi_n = \ln {1\over \bar R} + \left.
\sum_{k=6}^{n}\right.'' {1\over k-4}\>{z^{k-4}\over \bar R^{k-4}}\>
\eqno(5.13b)
$$
for $n\ge 6$ are the solution to (5.11).

The solutions (5.13) have an interesting property:
$$
{\partial \over \partial z}\Psi_n = - \>{z^{n-3}\over \bar R^{n-2}}\>,
\eqno(5.14)
$$
which will be used for finding point-like sources in Sec.6.

\subsection{Transformations from $(n-2)$D quantities to $n$D ones}

For anti-Laplacians there also exist a transformation converting
an $(n-2)$D solution to an $n$D one, mutual for odd-$n$D and even-$n$D
cases. The transformation acts starting from $n\ge 5$:
$$
\Psi_n = f_{A}^{(n,n-2)}\>\Psi_{n-2}= - \>{z^{n-3}\over n-4}\>
{\partial \over \partial z}\left( {1\over z^{n-4}}\> \Psi_{n-2}
\right).
\eqno(5.15)
$$
Using the differential rearrangement
$$
z^{n-3}\>{\partial \over \partial z}\>{1\over z^{n-4}}\>
\overline{\Delta}{}_z^{(n-2)} =
\overline{\Delta}{}_z^{(n)}\>
z^{n-3}\>{\partial \over \partial z}\>{1\over z^{n-4}}\>,
$$
we also find
$$
\overline{\Delta}{}_z^{(n)}\Psi_n = f_{A}^{(n,n-2)}
\overline{\Delta}{}_z^{(n-2)}\Psi_{n-2}.
\eqno(5.16)
$$
As before, we introduce the anti-$z$-Laplace set $A_n$:
$$
A_n = \{ \Psi_n,\>\overline{\Delta}{}_z^{(n)}\Psi_n,\>\bar s_z^{(n)}\}.
\eqno(5.17)
$$
where $\bar s_z^{(n)}$ is a point-like source searched for, which gives
rise to the solution $\Psi_n$:
$$
\overline{\Delta}{}_z^{(n)}\Psi_n =\bar s_z^{(n)},
$$
and whose explicit form we are just going to find using the
transformation
$$
\bar s_z^{(n)}=f_{A}^{(n,n-2)}\bar s_z^{(n-2)}.
\eqno(5.18)
$$
This is in contrast to Sec.4, where the validity of the transformation
$f_{L}^{(n,n-2)}$ acting on sources was merely checked.  This will be
done in the next sections separately for the odd-$n$D and even-$n$D
cases. In compact notations, the equalities (5.15), (5.16) and (5.18)
acquire the form
$$
A_n=f_{A}^{(n,n-2)}A_{n-2}.
$$

Any $n$D quantity from the anti-$z$-Laplace set (5.17) can be expressed
via the 3D or 4D one. For odd $n$,
$$
A_n = f_A^{(n,n-2)} f_A^{(n-2,n-4)}\ldots f_A^{(5,3)} A_3 =
{z^{n-3}\over (n-4)!!}
 \left(-\>{\partial \over \partial z}\> {1\over z}
\right)^{(n-3)/2} A_3,
\eqno(5.19)
$$
and for even $n$,
$$
A_n = f_A^{(n,n-2)}f_A^{(n-2,n-4)}\ldots f_A^{(6,4)} A_4 =
{z^{n-3}\over (n-4)!!} \left(-\>{\partial \over \partial z}\>
{1\over z} \right)^{(n-4)/2} {A_4\over z}\>.
\eqno(5.20)
$$

\section{Obtaining the point sources for potential-like
solutions to anti-$z$-Laplace equations}

\subsection{The simple odd-$n$D case}

In the odd-$n$D case, the 3D quantities of the Laplace and anti-Laplace sets
coincide: $A_3=L_3$. Indeed, $\Delta_{r,z}^{(3)}=\overline{\Delta}{}_z^{(3)}$,
$\Xi_3=\Psi_3$ and $s_{r,z}^{(3)}=\bar s_z^{(3)}$. This fact outlines several
ways to calculate $\bar s_z^{(n)}$. The first of them, the most simple and the
most formal is to apply formula (5.19) to $\bar s_z^{(3)}$, the result
being
$$
\bar s_z^{(n)} = - \left(\left.\sum_{k=3}^n \right.'a_{k,n}\right)
{\delta_+(r)\over r}\>\delta_+(z)= -\> {(n-3)!!\over (n-4)!!}
\>{\delta_+(r)\over r}\>\delta_+(z),
\eqno(6.1)
$$
see (B.3).
Perhaps, anybody could say that there should be taken more care in
dealing with formal expressions. That is why we give the second
(improved) way for obtaining the same result.

If to compare the expressions (4.20) and (5.19) for the transformations
of the Laplace and anti-Laplace sets, respectively, it is clear that the later differs
from the former by the presence of the factor $z^{n-3}$ and by the
permutation of $1/z$ and $\partial /\partial z$. This fact enables us
to express algebraically any quantity $A_n$ via $L_3, L_5, \ldots, L_{n-2},
L_n$. The expected expression,
$$
A_n = \left.\sum_{k=3}^n \right.' a_{k,n} z^{k-3}L_k,
\eqno(6.2)
$$
can be obtained by a recurrent way. Let us choose the second quantities from
the sets (4.18) and (5.17), then from (6.2)
$$
\overline{\Delta}{}_z^{(n)}\Psi_n = \left.\sum_{k=3}^n \right.' a_{k,n} z^{k-3}
 \Delta_{r,z}^{(k)} \Xi_k.
\eqno(6.3)
$$
Further, we multiply (6.3) on $r\phi$, where $\phi(r,z)$ belongs to the
above-mentioned class of trial functions, and perform integration with
respect to $r$ and $z$, cf. (4.29):
$$
\left< r\phi,\> \overline{\Delta}{}_z^{(n)}\Psi_n \right>
= \left.\sum_{k=3}^n \right.' a_{k,n}
\left< rz^{k-3}\phi,\> \Delta_{r,z}^{(k)} \Xi_k\right>.
\eqno(6.4)
$$
The $k$-th integrals on the r.h.s. of (6.4) were already calculated,
see (4.33) and (4.34), thus
$$
\left< r\phi,\> \overline{\Delta}{}_z^{(n)}\Psi_n \right>
= - \left(\left.\sum_{k=3}^n \right.'a_{k,n}\right)\phi(0,0),
\eqno(6.5)
$$
meaning the validity of (6.1).

At least, the third and the most accurate way is to examine the limit
of the (4.23)-type:
$$
\left< r\phi,\> \overline{\Delta}{}_z^{(n)}\Psi_n \right>
= \lim_{\varepsilon,\eta \to 0} (\bar J_r + \bar J_z)
$$
with the following surface terms obtained with the use of (6.3):
$$
\bar J_r = \left[\int\limits_0^{\varepsilon} dr\>r \left( \phi
\left.\sum_{k=3}^n \right.' a_{k,n} z^{k-3}\>{\partial \Xi_k\over \partial z}
- {\partial \phi\over \partial z}\>\Psi_n
\right) \right]_{z=\eta},
$$
$$
\bar J_z = \left[\int\limits_0^{\eta} dz \>r \left( \phi\>
{\partial \Psi_n\over \partial r} - {\partial \phi\over \partial r}\>\Psi_n
\right) \right]_{r=\varepsilon}.
$$
The result, being independent of the order of taking limits with respect to
$\varepsilon$ and $\eta$, certainly coincides with (6.5).

In order to avoid a sum factor on the r.h.s. of (6.1) or (6.5), it is
worth replacing the coefficients $a_{k,n}$ in (5.12) by the renormalized
ones $b_{k,n}$, see Appendix B and especially (B.5). Thus, our final
result is that for odd $n\ge 3$ the function
$$
\bar \Psi_n = \left.\sum_{k=3}^n \right.'
{(k-4)!!\>(n-k-1)!!\over (k-3)!!\>(n-k)!!}\>{z^{k-3}\over \bar R^{k-2}}
$$
satisfies the equation
$$
\overline{\Delta}{}_z^{(n)}\bar \Psi_n  \equiv \left(
{1\over r}\>{\partial \over \partial r}\> r\>{\partial \over \partial r}
+ z^{n-3}\> {\partial \over \partial z}\>
{1\over z^{n-3}}\>{\partial \over \partial z}\right)\bar \Psi_n
= - \>{\delta_+(r)\over r}\>\delta_+(z).
$$
Due to this renormalization, the $n$ dependence in the source on the
r.h.s. of the latter equation has disappeared.

\subsection{The difficult even-$n$D case}

For even $n$, the quantity $A_n$ cannot be reduced to any combination
of the quantities $L_k$.  Certainly, $A_2=L_2$ as a degenerate case,
however our transformation starts to work with $A_4\ne L_4$, cf. (4.21)
and (5.20). Fortunately, the solutions (5.13) possess the property
(5.14). Let us recall the relation (5.10) and impose $F(r,z)=\Psi_n$,
then combining equations (5.10),(5.14) and (4.13) gives
$$
{1\over z^{n-3}}\>{\partial \over \partial z}\>
\overline{\Delta}{}_z^{(n)}\Psi_n  = -\Delta_{r,z}^{(n)}
\left({1\over \bar R^{n-2}}\right)= \>{\delta_+(r)\over r}\>
 {\delta_+(z)\over z^{n-3}}.
\eqno(6.6)
$$
After multiplying (6.6) on $z^{n-3}$ and taking the antiderivative
with the use of the formal equality $\Theta_+'(z) = \delta_+(z)$,
$$
\overline{\Delta}{}_z^{(n)}\Psi_n  = \>{\delta_+(r)\over r}\>
\Theta_+(z) + \tilde f(r)
\eqno(6.7)
$$
where $\tilde f(r)$ is unknown as yet generalized function, and
$\Theta_+(z)$ is the Heaviside step function
$$
\Theta_+(x) =\cases{ 0, & for $x\le 0$;\cr
                  1,  & for $x>0$.\cr}
$$
The equality (6.7) suggests us to explicitly calculate the case $n=4$
with the $\Theta$ source.

Let us find the function $\tilde \Psi_4(r,z)$ which satisfies
the equation
$$
\overline{\Delta}{}_z^{(4)}\tilde \Psi_4  = \>{\delta_+(r)\over r}\>
\Theta_+(z).
\eqno(6.8)
$$
Combining (6.8) with (5.6) for $n=4$ and  with $F(r,z)$ replaced by
$\tilde \Psi_4(r,z)$, and redenoting $\theta_1=\theta$ lead to the
equation
$$
{\Delta}_{r,z,\theta}^{(4)}\left(
{\cos \theta\over z}\> \tilde \Psi_4(r,z)\right)
={\cos \theta\over z}\> \overline{\Delta}{}_z^{(4)}\tilde \Psi_4(r,z)
= {\cos \theta\over z}\>{\delta_+(r)\over r}\>\Theta_+(z)
\equiv - \sigma_4 \mu(r,z,\theta).
\eqno(6.9)
$$

The fundamental solution to the Poisson equation (4.1) in the case $n=4$
is
$$
\Phi_4 = {1\over 2|\vec R - \vec R'|^2}
$$
where
$$
|\vec R - \vec R'|^2 = r^2 + r'^2 -
2r'r\cos (\varphi-\varphi') + z^2 + z'^2 - 2z'z \cos (\theta-\theta').
$$
Thus, similarly to a derivation done in Sec.2 and in accordance with a
standard rule, the solution to (6.8) can be written as follows
\footnote{Strictly speaking, the standard rule prescribes to deal with
compact sources, although the source in (6.9) is not compact. Nevertheless,
the integral (6.10) exists. For accuracy, we could make a limit procedure
originating from integration over a confined space domain, and then
coming to the whole space.}
$$
{\cos \theta\over z}\> \tilde \Psi_4 = {1\over 2}\int_{\tau^{(4)}{}'}
d\tau^{(4)}{}'\>\>{\mu(r',z',\theta')\over |\vec R - \vec R'|^2}=
-\>{1\over 2\sigma_4}\int\limits_0^{\infty} dr'\int\limits_0^{2\pi} d\varphi'
\int\limits_0^{\infty}dz' \int\limits_0^{2\pi} d\theta' \times
$$
$$
{\cos \theta'\delta_+(r')\Theta_+(z')\over r^2+r'^2-2r'r\cos(\varphi-\varphi')
+z^2+z'^2-2z'z\cos(\theta-\theta')}\>.
\eqno(6.10)
$$
For integration of (6.10) with respect to angles, we require the following
handbook integrals (cf. \cite{Pr-Br-M}, p.181, no.5 and p.414, no.22):
$$
\int\limits_0^{\pi}{dx\over a+b\cos x} = {\pi\over \sqrt{a^2-b^2}}\>,
\eqno(6.11)
$$
$$
\int\limits_0^{\pi}{dx\>\cos x\over a+b\cos x} = {\pi\over b}\left(1-
{a\over \sqrt{a^2-b^2}}\right).
\eqno(6.12)
$$
Now we shall perform the three subsequent operations: 1) integration
with respect to $\varphi'$ with the use of (6.11), 2) integration with
respect to $r'$, and 3) the procedure of removing the factor
$\cos\theta$ after imposing $\theta'=\theta + \beta$ which was
already described in Sec.2. This leads to
$$
\tilde \Psi_4 =-\>{1\over \pi}\>z\int\limits_0^{\infty} dz'
\int\limits_0^{\pi} {d\beta\>\cos\beta \over r^2+z^2+z'^2-2z'z\cos\beta}\>.
\eqno(6.13)
$$
The next step is the integration of (6.13) with respect to $\beta$
using (6.12):
$$
\tilde \Psi_4 = -\>{1\over 2}\>\int\limits_0^{\infty} {dz'\over z'}
\left( {r^2+z^2+z'^2\over [(r^2+z^2)^2
+2(r^2-z^2)z'^2+z'^4]^{1/2}} - 1\right).
\eqno(6.14)
$$
The integral (6.14) can be calculated with the use of the following
handbook one (see loc. cit, p.102, no.8):
$$
\int {dx\over \sqrt{ax^2+bx+c)}} = {1\over \sqrt{a}}\>\ln\left|
{2ax+b\over 2\sqrt{a}} + \sqrt{ax^2+bx+c)}\right|.
$$
This final integration gives
$$
\tilde \Psi_4 = \ln {1\over \sqrt{r^2+z^2}} - \ln {1\over r}\>.
$$

Now, we have obtained the solution corresponding to the $\Theta$ source
(6.8). It differs from the announced solution to the homogeneous equation
by the only last term. However, the later is the degenerate 2D
solution (4.14b) taken with an opposite sign, and that is why it is
simultaneously the $z$ independent solution to the equation [cf. (4.13b)]:
$$
\overline{\Delta}{}_z^{(4)}\Xi_2 = \Delta^{(2)}\Xi_2 =
 {1\over r}\> {\partial \over \partial r}\>r\>
{\partial \over \partial r}\> \Xi_2 = - \>{\delta_+(r)\over r}\>.
$$
Taking the sum of the solutions  $\tilde \Psi_4$ and $\Xi_2$, we can state
that the function (5.13a),
$$
\Psi_4= \tilde \Psi_4 + \Xi_2 = \ln {1\over \bar R}\>,
\eqno(6.15)
$$
satisfies the equation with a point-like source
$$
\overline{\Delta}{}_z^{(4)}\Psi_4 =  - \>{\delta_+(r)\over r}\>
[1-\Theta_+(z)] \equiv \bar s_z^{(4)}.
\eqno(6.16)
$$

Indeed, for the $z$ range of values: $z\ge 0$,
$$
1-\Theta_+(z) = \cases{ 1, & for $z=0$;\cr
                        0, & for $z>0$.\cr}
\eqno(6.17)
$$
We should remark that the situation is somewhat paradoxical. The
function (6.17) is finite (nonsingular) in the point $z=0$; and the
source in (6.16) although located at the point $r=0,z=0$ is more "weak"
than the true $\delta$ source. However, the logarithmic divergency in
(6.15) when reaching zero is more weak than the divergency $\sim \bar
R^{-2}$ as it could be in the case $n=4$ for the true $\delta$ source.
If we tried to extract this source by the standard method involving
trial functions as it was done before, we would obtain zero when
integrating over the whole $4$D space, although the derivative of
(6.17) with respect to $z$ is the $-\delta_+$ function as usual.

There exists a theorem stating that every generalized function
concentrated at a point is a combination of the $\delta$ functions and
their derivatives \cite{G-Sh-2}, the proof being done in terms of
finite functionals on continuous functions. The theorem has a
consequence that a solution to the Laplace equation with a power-law
singularity is generated by this combination. It is also extended onto
partial differential equations with constant coefficients (in a certain
frame, if any). Our situation is more sophisticated. The author's
opinion is that, first, there are no finite nonzero functional
determined by the generalized function (6.17) in any class of trial
(continuous) functions. Second, the equations with anti-Laplacians for
every $n>3$ and the even-$n$D solutions considered are not the
equations and solutions of the above type. Certainly, additional
investigations to this situation from the viewpoint of functional
analysis are required.

In order to extend the obtained $4$D source on subsequent even $n$, we
can certainly use the transformation $f_{A}^{(n,n-2)}$. One could easily
ensure that it does not change the form of the source,
$$
f_{A}^{(n,n-2)}\>[1-\Theta_+(z)] = - \>
{z^{n-3}\over n-4}\> {\partial \over \partial z}
\left({1\over z^{n-4}}\>[1-\Theta_+(z)]\right) =
1-\Theta_+(z),
$$
due to the formal equality $z\delta_+(z)=0$. Our previous remark can be
transferred mutatis mutandis to the $n$D case as well.

Thus, we establish that for even $n\ge 6$ the function (5.13b),
$$
\Psi_n = \ln {1\over \bar R} + \left.
\sum_{k=6}^{n}\right.'' {1\over k-4}\>{z^{k-4}\over \bar R^{k-4}}\>,
$$
is the solution to the equation
$$
\overline{\Delta}{}_z^{(n)}\Psi_n =  - \>{\delta(r)_+\over r}\>
[1-\Theta_+(z)] \equiv \bar s_z^{(n)}.
$$
Fortunately, we have chose from the beginning such a mutual coefficient
at $\Psi_n$ that $\bar s_z^{(n)}$ has no dependence on $n$.

\section{Returning to the vortex-ring-like solutions. Conclusion}

As before, introduce the anti-double-Laplacian set $D_n$,
$$
D_n = \{ \psi_n,\>\overline{\overline{\Delta}}{}^{(n)}\psi_n,\>
\overline{\overline{s}}{}^{(n)}\}.
$$
Given such a powerful tool as the transformation $f_{A}^{(n,n-2)}$ in
(5.15), we have no problems in constructing any quantity $D_n$ from
$D_{n-2}$. Indeed, the distinction between corresponding $A_n$ and
$D_n$ is entirely referred to their "$r$ parts", the "$z$ parts" being
unchanged. Without such an explanation, it can be immediately verified
that
$$
D_n=f_{A}^{(n,n-2)}D_{n-2}.
$$
Expected results can be formulated as before separately for the two cases.

In the odd-$n$D case, we should start with $D_3$: As to the operators,
$\overline{\overline{\Delta}}{}^{(3)}$ coincides with
$\overline{\Delta}{}_r^{(3)}$, and the latter is the redenoted operator
(2.4). Unlike Sec.2, where we have defined $\Omega_\varphi$ by a
traditional way, here we make another definition,
$$
\int\limits_0^{\infty}dr\>r \int\limits_0^{2\pi}d\varphi
\int\limits_{-\infty}^{\infty}dz\>\>\Omega_\varphi = 4\pi,
\eqno(7.1)
$$
in order for the source $\overline{\overline{s}}{}^{(3)}$ to have the form
$$
\overline{\overline{s}}{}^{(3)}=-\delta (r-a) \delta_+ (z)
\eqno(7.2)
$$
instead of (2.7). [We have imposed $\delta(z)=\delta(\vec z)=\delta_+(z)/2$
in (7.2) according to the general formula (4.3).] Hence, the 3D vortex-ring
solution can be rewritten in the form
$$
\psi_3 = {1\over \pi}\>r \int\limits_0^{\pi} d\alpha \cos \alpha\>
{1\over R}\>,
$$
instead of (2.11), recall from (3.7) that
$$
R = [r^2 + a^2 - 2ar\cos\alpha + z^2]^{1/2}.
$$

The final result is that, for odd $n\ge 3$, the function
$$
\psi_n = {1\over \pi}\>r \left.\sum_{k=3}^n \right.'
{(k-4)!!\>(n-k-1)!!\over (k-3)!!\>(n-k)!!}
\int\limits_0^{\pi} d\alpha \cos\alpha\>
{z^{k-3}\over R^{k-2}}
$$
satisfies the equation
$$
\overline{\overline{\Delta}}{}^{(n)}\psi_n
\equiv \left(r\>{\partial \over \partial r}\>{1\over r}\>{\partial \over \partial r}
+ z^{n-3}\> {\partial \over \partial z}\>
{1\over z^{n-3}}\>{\partial \over \partial z}\right)\psi_n
= - \delta(r-a)\>\delta_+(z) \equiv \overline{\overline{s}}{}^{(n)}.
$$

In the even-$n$D case, the same problem as that in the subsection 6.2
arises, however, it can be resolved by one-to-one correspondence at
each step of a derivation. We only recall that instead of the solution
(4.13b) to equation (4.14b), we have to use the solution (3.25) to
equation (3.24) with a source $\propto \delta(r-a)$. Thus, for $n=4$
and even $n\ge 6$, the functions
$$
\psi_4 = {1\over \pi}\>r\int\limits_0^{\pi} d\alpha \cos\alpha\>
\ln {1\over R}\>
\eqno(7.3)
$$
and
$$
\psi_n = {1\over \pi}\>r\int\limits_0^{\pi} d\alpha \cos\alpha\>
\ln {1\over R}\> + {1\over \pi}\>r\left.\sum_{k=6}^{n}\right.''
{1\over k-4}\int\limits_0^{\pi} d\alpha \cos\alpha\>
{z^{k-4}\over R^{k-4}}\>,
$$
respectively, are the solutions to the equation
$$
\overline{\overline{\Delta}}{}^{(n)}\psi_n = -\delta(r-a)\>[1-\Theta_+(z)]
\equiv \overline{\overline{s}}{}^{(n)}.
$$

Now, several concluding remarks are in order.

{\bf 1.}
Certainly, every $n$D solution to equations involving the
anti-$z$-Laplacian and anti-double-Laplacian can in principle be
obtained by the procedure described in Sec.2, namely, by using the
relations (5.6) and (5.7) and by integrating a fundamental solution to
the Poisson equation multiplied by a corresponding source. However,
such the way gives rise to great practical difficulties. In the $(r,z)$
frame, an immediate integration is still relatively easy for $n=4$, as
it was demonstrated in the subsection 6.2, but even for $n=5$ an
integrand arises which contains special functions with arguments
involving trigonometric functions.  These difficulties grows when
coming to each subsequent $n$. That is why the transformation
$f_{A}^{(n,n-2)}$ saves the situation:  There are no problems in
obtaining solutions with an arbitrary (large) $n$.

{\bf 2.}
All the obtained here solutions have one more advantage that they are
some algebraic functions of $z$ and $\bar R$ (or $R$) plus the
logarithmic function of $\bar R$ (or $R$) for even $n$. In the latter
case both the anti-$z$-Laplace and anti-double-Laplace solutions
corresponding to sources with $\delta_+(z)$ instead of $1-\Theta_+(z)$
are also feasible. However, they involve inverse trigonometric
functions of $r$ and $z$, and have no such remarkable announced
properties. By virtue of the relation (5.9b), the odd-$n$ solutions
related to a $\delta_+'(z)$ are also feasible, but this subject is
outside the framework of our study.

{\bf 3.}
Is was technically more simple to work with the anti-$z$-Laplacians
than with the anti-double-Laplacians. It was also more simple to
extract the form of point sources than that of ring sources.  Moreover,
this study has clarified the fact that the solutions corresponding to
both the operators are of a similar type -- it does not matter whether
their sources are ring-like or point-like ones.

However, there also exist ring-like solutions corresponding to the
anti-$z$-Laplacians.  We add them to the list of previous solutions.
They are "scalar" in their "$r$ parts" without changes in the "$z$
parts", so that the transformation $f_{A}^{(n,n-2)}$ is working as
before. As the result, for odd $n$, the function
$$
\chi_n ={1\over \pi}\>\left.\sum_{k=3}^n \right.'{(k-4)!!\>(n-k-1)!!\over
(k-3)!!\>(n-k)!!}\int\limits_0^{\pi} d\alpha \>
{z^{k-3}\over R^{k-2}}
\eqno(7.4)
$$
$(n\ge 3)$ satisfies the equation
$$
\overline{\Delta}{}_z^{(n)}\chi_n = - {\delta(r-a)\over r}\>\delta_+(z).
$$
While for $n=4$ and even $n\ge 6$, the functions
$$
\chi_4 = {1\over \pi}\int\limits_0^{\pi} d\alpha \>
\ln {1\over R}\>
\eqno(7.5)
$$
and
$$
\chi_n = {1\over \pi}\int\limits_0^{\pi} d\alpha \>
\ln {1\over R}\> + {1\over \pi}\>\left.\sum_{k=6}^{n}\right.''
{1\over k-4}\int\limits_0^{\pi} d\alpha \>
{z^{k-4}\over R^{k-4}}\>,
\eqno(7.6)
$$
respectively, are the solutions to the equation
$$
\overline{\Delta}{}_z^{(n)}\chi_n = -{\delta(r-a)\over r}\>[1-\Theta_+(z)].
$$
In both the cases, the integral identity
$$
\int\limits_0^{\pi} d\alpha \left({\cos\alpha\over R^k} - k\>{ar\sin^2\alpha
\over R^{k+2}}
\right)= \int\limits_0^{\pi} d\alpha\>{\partial \over \partial \alpha}
\left({\cos\alpha\over R^k}
\right)= 0
$$
is necessary if one desires to immediately verify that the functions (7.4),
(7.5) and (7.6) satisfy the homogeneous equations
$\overline{\Delta}{}_z^{(n)} \chi_n = 0$.

{\bf 4.} Until now we have said nothing about solutions corresponding
to the $(n>3)$-dimensional anti-$r$-Laplacians (5.3). The extension of
equation (2.4) onto the $n$D space is
$$
\overline{\Delta}_r^{(n)} \xi_n \equiv \left(r\>{\partial \over \partial r}\>
{1\over r}\>{\partial \over \partial r} + {1\over z^{n-3}}\>
{\partial \over \partial z}\>z^{n-3}\>{\partial \over \partial z}\right)\xi_n
= -r \Omega_\varphi^{(n)} (r,z) \equiv \bar s_r^{(n)}.
\eqno(7.7)
$$
We define $\Omega_\varphi^{(n)}$ by a way which generalizes (7.1)
to the $n$D case:
$$
\int_{\tau^{(n)}} d\tau^{(n)}\>\Omega_\varphi^{(n)} = (n-2)\sigma_n
\eqno(7.8)
$$
with $d\tau^{(n)}$ determined by (A.4). (Here, both the odd-$n$D and
even-$n$D cases can be considered in common.) After that the source in
(7.7) acquires the form
$$
\bar s_r^{(n)} = - \sigma_{n-2}\delta(r-a)\>\delta(\vec z)
= - \delta(r-a)\>{\delta_+(z)\over z^{n-3}}.
$$
This case is very simple. It can be resolved using the relation (5.5)
and dealing with the Cartesian frame in the $z$ space where
$\delta(\vec z)=\delta(z_1)\delta(z_2)\ldots\delta(z_{n-2})$, see
Appendix A. The answer is that the function
$$
\xi_n = {1\over \pi}\>r\int\limits_0^{\pi} d\alpha \cos \alpha \>
{1\over R^{n-2}}
$$
for all $n\ge 3$ is the solution to the equation
$$
\overline{\Delta}_r^{(n)} \xi_n =- \delta(r-a)\>{\delta_+(z)\over z^{n-3}}\>.
$$
We see that this solution has no attractive properties of those
considered before. It exhibits a typical $n$D-space behaviour.
Nevertheless, due to the choice of (7.8), the anti-$r$-Laplacian set $B_n$:
$$
B_n = \{\xi_n,\> \overline{\Delta}_r^{(n)}\xi_n,\>\bar s_r^{(n)}\},
$$
also has the property of the type (4.19):
$$
B_n = f_L^{(n,n-2)} B_{n-2}
$$
where the transformation is involved which is suitable for the
Laplace set.

{\bf 5.} Emphasize one more that coming from a "Laplacian part" of an
operator to an "anti-Laplacian part" for a given subspace means coming
from scalars to other geometrical objects. Now it is already clear, by
analogy with Sec.2, that in the case $n=4$ the presented solutions,
i.e. the functions (6.15) and (7.3), are the $\theta$ components of
vectors in the 2-dimensional $z$ space orthogonal to the radial
direction. For larger $n$, the situation seems to be more intricate
because components of some polyvector are dealth with. The author is
going to devote to this subject further studies.

\subsection*{Appendix A. The $x$ and $(r,z)$ frames}

In $n$ dimensions, the spherical coordinate frame is given by the radial
distance and the $n-1$ angles, $x$ and $\lambda_1,\lambda_2,\ldots,
\lambda_{n-1}$ in our notations, respectively. We call this frame the
$x$ frame and $\lambda_1$ the senior angle. Note that the
angle ranges of values are $0\le\lambda_k\le\pi$ for $k=1,2,\ldots,n-2$
and $0\le\lambda_{n-1}\le 2\pi$.

In  transforming the $x$ frame into a Cartesian one, we introduce the
notations
$$
\begin{array}{rcl}
x_1&=&x\cos \lambda_1,\\
x_2&=&x\sin \lambda_1 \cos \lambda_2,\\
   & & \ldots \\
x_{n-1}&=&x\sin \lambda_1 \sin \lambda_2 \ldots \sin \lambda_{n-1}
\cos \lambda_{n-1},\\
x_n&=&x\sin \lambda_1 \sin \lambda_2 \ldots \sin \lambda_{n-1}
\sin \lambda_{n-1}.\\
\end{array}
$$
The most compact expression for metric can be written as follows
$$
dl^2= g_{\mu \nu}dx^{\mu}dx^{\nu}=dx^2+x^2 d\varsigma_1^2
\eqno({\rm A}.1)
$$
($\mu,\nu = 1,2,\ldots,n$) where
$$
\begin{array}{rcl}
d\varsigma_1^2 &=& d\lambda_1^2 + {\sin}^2 \lambda_1 d\varsigma_2^2,\\
d\varsigma_2^2 &=& d\lambda_2^2 + {\sin}^2\lambda_2  d\varsigma_3^2,\\
&& \ldots \\
d\varsigma_{n-2}^2 &=& d\lambda_{n-2}^2 + {\sin}^2\lambda_{n-2}^2
d\varsigma_{n-1}^2\\
d\varsigma_{n-1}^2 &=& d\lambda_{n-1}^2.\\
\end{array}
$$
In other words,
$$
\begin{array}{rcl}
g_{xx}&=&1,\\
g_{\lambda_1\lambda_1}&=&x^2,\\
g_{\lambda_2\lambda_2}&=&x^2 {\sin}^2 \lambda_1,\\
&& \ldots \\
g_{\lambda_{n-1}\lambda_{n-1}}&=&x^2 {\sin}^2 \lambda_1
{\sin}^2 \lambda_2 \ldots {\sin}^2 \lambda_{n-2},\\
\end{array}
$$
The covariant volume element for the metric (A.1) is
$$
d\tau^{(n)}= d\sigma_n x^{n-1} dx
$$
with
$$
d\sigma_n={\sin}^{n-2} \lambda_1
{\sin}^{n-3} \lambda_2 \cdots \sin \lambda_{n-2}\>d\lambda_1\>d\lambda_2
\cdots d\lambda_{n-2}\>d\lambda_{n-1},
$$
note that $\int d\sigma_n=\sigma_n$ defined by (4.2).

The scalar Laplacian for the metric (A.1) has the form
$$
\Delta_x^{(n)}={1\over x^{n-1}}\> {\partial \over \partial x}\> x^{n-1} \>
{\partial \over \partial x}+{1\over x^2}\left(
{1\over \sin^{n-2}\lambda_1}\>{\partial \over \partial \lambda_1}\>
\sin^{n-2}\lambda_1\>{\partial \over \partial \lambda_1}
+ {1\over \sin^2\lambda_1 \sin^{n-3}\lambda_2}\>\times \right.
$$
$$
\left. {\partial \over \partial \lambda_2}\>
\sin^{n-3}\lambda_2\>{\partial \over \partial \lambda_2} + \ldots
+ {1\over \sin^2\lambda_1 \sin^2\lambda_2\cdots\sin^2\lambda_{n-2}\sin
\lambda_{n-2}}\>{\partial^2 \over \partial \lambda_{n-1}^2}
\right).
\eqno({\rm A}.2)
$$
\vskip0.5cm

We introduce the $(r,z)$ frame which is the direct product of a
2-dimensional and $(n-2)$-dimensional spherical frames with the
coordinates $r,\varphi$ and $z,\theta_1,\theta_2$, $\ldots$,
$\theta_{n-3}$, respectively. All the above quantities can be rewritten
mutatis mutandis.  We take a Cartesian frame as a direct product of the
2-dimensional and $(n-2)$-dimensional Cartesian frames with the
coordinates
$$
\begin{array}{rcl}
r_1&=&r\cos \varphi,\\
r_2&=&r\sin \varphi,\\
z_1&=&z\cos \theta_1,\\
z_2&=&z\sin \theta_1 \cos \theta_2,\\
&&\ldots\\
z_{n-3}&=&\sin \theta_1\sin \theta_2\ldots\sin \theta_{n-4}\cos \theta_{n-3},\\
z_{n-2}&=&\sin \theta_1\sin \theta_2\ldots\sin \theta_{n-4}\sin \theta_{n-3}.\\
\end{array}
$$
In contrast to (A.1), we give below another (noncompact) expression for
metric:
$$
dl^2=dr^2+r^2d\varphi^2+dz^2+z^2(d\theta_1^2+\sin^2\theta_1(d\theta_2^2+
\sin^2\theta_2(\ldots + \sin^2\theta_{n-2}d\theta_{n-3}^2)\ldots )).
\eqno({\rm A}.3)
$$
The volume element for the metric (A.3) is
$$
d\tau^{(n)}= d\sigma_2 d\sigma_{n-2} rz^{n-3}dr\>dz
\eqno({\rm A}.4)
$$
with
$$
\begin{array}{rcl}
d\sigma_2&=&d\varphi,\\
d\sigma_{n-2}&=&\sin^{n-4}\theta_1\sin^{n-5} \theta_2\cdots\sin \theta_{n-4}
\>d\theta_1\>d\theta_2\cdots d\theta_{n-4}\>d\theta_{n-3}.\\
\end{array}
$$

The divergence and the scalar Laplacian are respectively as follows,
$$
\vec \nabla \cdot \vec v = {1\over \sqrt{g}}\>{\partial \over \partial
x^{\mu}}\> (\sqrt{g} v^{\mu}) =
{1\over r}\> {\partial \over \partial r}\>(rv^r)+{1\over r^2}\>
{\partial v^{\varphi}\over \partial \varphi} +
{1\over z^{n-3}}\> {\partial \over \partial z}\>(z^{n-3}v^z)\> +
$$
$$
{1\over \sin^{n-4}\theta_1}\>{\partial \over \partial \theta_1}\>
(\sin^{n-4}\theta_1 v^{\theta_1}) +
{1\over \sin^{n-5}\theta_2}\>{\partial \over \partial \theta_2}\>
(\sin^{n-5}\theta_2 v^{\theta_2})
$$
$$
 + \ldots +
{1\over \sin\theta_{n-4}}\>{\partial \over \partial \theta_{n-4}}\>
(\sin\theta_{n-4} v^{\theta_{n-4}}) +
{\partial v^{\theta_{n-3}}\over \partial \theta_{n-3}}
\eqno({\rm A}.5)
$$
where $g=\det g_{\mu\nu}$, and
$$
\Delta_{r,z}^{(n)}={1\over r}\> {\partial \over \partial r}\>r\>
{\partial \over \partial r} + {1\over r^2}\>
{\partial^2 \over \partial \varphi^2} + {1\over z^{n-3}}\>
{\partial \over \partial z}\> z^{n-3}\>{\partial \over \partial z}\>+
$$
$$
{1\over z^2}\left(
{1\over \sin^{n-4}\theta_1}\>{\partial \over \partial \theta_1}\>
\sin^{n-4}\theta_1\>{\partial \over \partial \theta_1}
+ {1\over \sin^2\theta_1 \sin^{n-5}\theta_2}\>
{\partial \over \partial \theta_2}\>
\sin^{n-5}\theta_2\>{\partial \over \partial \theta_2}
\right.
$$
$$
\left. + \ldots +
{1\over \sin^2\theta_1 \sin^2\theta_2 \cdots
\sin^2\theta_{n-5}\sin\theta_{n-4}}\>
{\partial^2 \over \partial \theta_{n-3}^2}
\right).
\eqno({\rm A}.6)
$$

\section*{Appendix B. More about $a_{k,n}$}

The coefficients $a_{k,n}$ can be represented in the form equivalent to
(3.13):
$$
a_{k,n}={(n-3)!!\over (n-4)!!}\>{(k-4)!!\>(n-k-1)!!\over (k-3)!!\>(n-k)!!}\>.
\eqno({\rm B}.1)
$$
Note that $(-1)!!=1$ and $0!!=1$ by definition. It is quite clear from
(B.1), that these coefficients possess the symmetry property
$$
a_{k,n}=a_{n-k+3,n}.
\eqno({\rm B}.2)
$$
In particular,
$$
a_{n,n}=a_{3,n}=1.
$$
Besides the recurrence relation (3.12), there exists
one more recurrence relation for $k\le n-2$:
$$
a_{k,n}=a_{k,n-2}\>{(n-3)(n-k-1)\over (n-4)(n-k)}\>.
$$
The above formula permits one to rapidly calculate the triangle of the
numerical values of $a_{k,n}$. We give it below for $n=3,5,7,9,11,13,15$.
$$
\begin{array}{lllccccccccccccccccccc}
n=3&&&&&&1\\
n=5&&&&&&1&1\\
n=7&&&&&&1&2/3&1\\
n=9&&&&&&1&3/5&3/5&1\\
n=11&&&&&&1&4/7&18/35&4/7&1\\
n=13&&&&&&1&5/9&10/21&10/21&5/9&1\\
n=15&&&&&&1&6/11&5/11&100/231&5/11&6/11&1\\
\ldots&&&&&&&&&\ldots\\
\end{array}
$$

By immediate calculations for several odd $n$, we can check the three
following heuristical relations. The first of them is the relation
for the required sum,
$$
\left.\sum_{k=3}^n \right.' a_{k,n} = {(n-3)!!\over (n-4)!!}\>,
\eqno({\rm B}.3)
$$
and the two other are the interesting although useless relations
($n>3$)
$$
\left.\sum_{k=3}^n \right.' {a_{k,n}\over k-4}=0, \qquad
\left.\sum_{k=3}^n \right.' {a_{k,n}\over n-k-1}=0,
$$
the latter follows from the former when using (B.2).

In order to surely use (B.3), we must prove it. To do this, let us
renormalize the coefficients $a_{k,n}$:
$$
a_{k,n}={(n-3)!!\over (n-4)!!}\>b_{k,n},
\eqno({\rm B}.4)
$$
$$
b_{k,n}={(k-4)!!\>(n-k-1)!!\over (k-3)!!\>(n-k)!!}\>.
\eqno({\rm B}.5)
$$
According to (B.3) and (B.4), we must prove that
$$
\left.\sum_{k=3}^n \right.' b_{k,n} = 1.
$$
Let us impose $k=2l+1$ and $n=2q+1$ and use the equality
$$
{-1/2\choose m} = (-1)^m\>{(2m-1)!!\over (2m)!!}\>,
$$
see \cite{Pr-Br-M}, p.772, where by definition
$$
{a\choose b} = {\Gamma(a+1)\over \Gamma(b+1)\Gamma(a-b+1)}\>.
$$
The coefficients (B.5) can be now expressed as follows
$$
b_{k,n}=\tilde b_{l,q}= (-1)^{q-1} {-1/2\choose l-1} {-1/2\choose q-l}.
$$
Let us write the expression for the following sum
$$
\sum_{l=1}^q {-1/2\choose l-1} {-1/2+\epsilon\choose q-l}=
{-1+\epsilon\choose q-1}
\eqno({\rm B}.6)
$$
in accordance with the handbook equality for any complex $a$ and $b$
(see loc. cit., no.13 on p.616):
$$
\sum_{k=0}^n {a\choose k} {b\choose n-k}=
{a+b\choose n}.
$$
It is necessary to obtain the sum (B.6) for $\epsilon=0$.
However, in this case (B.6) contains an indeterminate
form on its r.h.s., in order to evaluate it we take the limit
$\epsilon\to 0$:
$$
\eqalign{
 \lim_{\epsilon \to 0}{-1+\epsilon\choose q-1} & =
\lim_{\epsilon \to 0}{\Gamma(\epsilon)\over \Gamma(q)\Gamma(1-q+\epsilon)}\cr
& = \lim_{\epsilon \to 0}
{\Gamma(\epsilon)\over \Gamma(q-\epsilon)\Gamma(1-q+\epsilon)}=
\lim_{\epsilon \to 0} {\sin \pi(q-\epsilon)\over \epsilon\>\pi}=(-1)^{q-1}\cr
}
$$
where we have replaced the finite value $\Gamma(q)$ on $\Gamma(q-\epsilon)$
under the sign of limit. At least,
$$
\left.\sum_{k=3}^n \right.' b_{k,n}  = \sum_{l=1}^q \tilde b_{l,q} =
(-1)^{q-1}\lim_{\epsilon \to 0}{-1+\epsilon\choose q-1} =1,
$$
that finishes our proof. Perhaps, a more elegant proof is available,
however, the author could not find it.

We think it is worth giving the triangle of the coefficients $b_{k,n}$
(similar to that of $a_{k,n}$) because our final expressions for
odd-$n$D solutions contain just $b_{k,n}$.
$$
\begin{array}{lllccccccccccccccccccc}
n=3&&  1\\
n=5&&  1/2 & 1/2\\
n=7&&  3/8 & 1/4 & 3/8\\
n=9&&  5/16 & 3/16 & 3/16 & 5/16\\
n=11&& 35/128 & 5/32 & 9/64 & 5/32 & 35/128\\
n=13&& 63/256 & 35/256 & 15/128 & 15/128 & 35/256 & 63/256\\
n=15&& 231/1024 & 63/512 & 105/1024 & 25/256 & 105/1024 & 63/512 & 231/1024\\
\ldots&&&&&\ldots\\
\end{array}
$$

\end{document}